\ifx\pdfoutput\undefined
	\documentclass[dvips]{cimento}
\else 
  \documentclass[pdftex]{cimento} 
\fi 

\usepackage{graphicx}
\usepackage{amsmath}  
\usepackage{amssymb}  
\usepackage{latexsym}  

\def\rmd{\rm d}
\def\beq{\begin{equation}}
\def\eeq{\end{equation}}

\title{Charges in gravitational fields:
from Fermi, via Hanni-Ruffini-Wheeler, to the \lq\lq electric Meissner effect''}
\author{
R. Ruffini\from{ins:roma1}\from{ins:icra}\thanks{ruffini@icra.it}
}
\instlist{
\inst{ins:roma1} Dipartimento di Fisica, Universit\`a ``La Sapienza'' di Roma, I--00185, Italy  
\inst{ins:icra} International Center for Relativistic Astrophysics - I.C.R.A., University of Rome ``La Sapienza'', I--00185 Rome, Italy
}
\PACSes{\PACSit{04.20}{}
\PACSit{04.20}{}}
\begin{document}

\maketitle

\begin{abstract}
Recent developments in obtaining a detailed model for gamma ray bursts have shown the need for a deeper understanding of phenomena described by solutions of the Einstein-Maxwell equations, reviving interest in the behavior of charges close to a black hole. In particular a drastic difference has been found between the lines of force of a charged test particle in the fields of Schwarzschild and Reissner-Nordstr\"om black holes. This difference characterizes a general relativistic effect for the electric field of a charged test particle around a (charged) Reissner-Nordstr\"om black hole similar to the ``Meissner effect'' for a magnetic field around a superconductor. These new results are related to earlier work by Fermi and Hanni-Ruffini-Wheeler.
\end{abstract}

\section{Introduction}

It is a great pleasure to celebrate this eightieth birthday of M.me Choquet-Bruhat.
M.me Choquet-Bruhat has collaborated with us in organizing the Marcel Grossman Meeting series from many years,
as did Yakov Borisovich Zel'dovich, who had also been a long standing member of the International Organization Committee of the Grossman Meetings as well as a good friend of both of us. I would like to begin with an anecdote about him.
The achievements of Zel'dovich are well known worldwide: he had invented the Katiuscia Rocket which had played an essential role in the history of the Second World War. 
He then developed both the atomic and thermonuclear bombs of the Soviet Union with Zakharov and was also instrumental in the development of space research in the Soviet Union. 
It was not until 1960 that Zel'dovich became interested in relativistic astrophysics and developed his internationally recognized school of research on this topic in the Soviet Union.

A great scientist may contribute to the progress of science not only by his rational scientific works, but also by his mental extravagance. In this sense Zel'dovich triggered one of the greatest discoveries ever in relativistic astrophysics: the gamma ray bursts (GRBs). For the understanding of these sources it is essential to understand the process of energy extraction from a black hole, an energy we have called blackholic energy. In turn this theoretical research on the vacuum polarization process in the field of a black hole has demanded a deeper understanding of the interrelationships between Maxwell's equations and the Einstein equations. Exactly this problematic convinced us of the need to go back to some of the classic works on the interaction between a charged test particle and a black hole. It has been very fortunate that from this analysis  new aspects of physics have surfaced which will be briefly summarized in this talk. I would also like to emphasize that this work finds its origin in a paper by Enrico Fermi which is largely unknown and published only in Italian, only being translated into English this year \cite{fermibook}.   

\begin{figure} 
\centering 
\includegraphics[width=\hsize,clip]{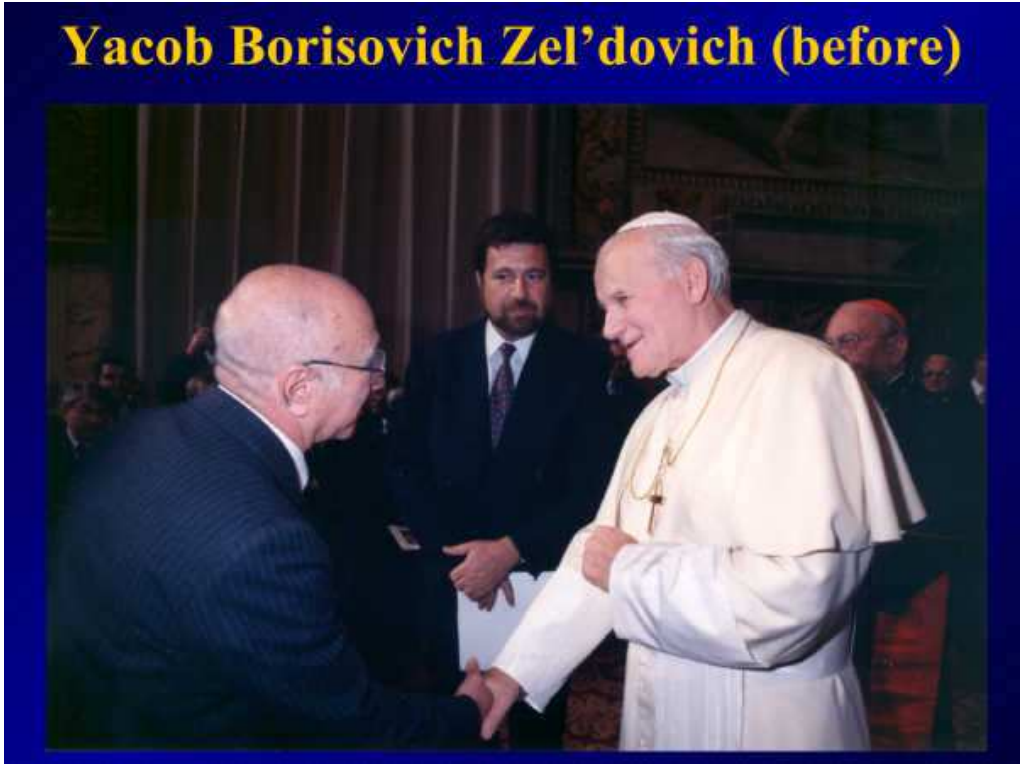} 
\caption{Y.B. Zel'dovich being introduced to the Pope John Paul II by Remo Ruffini (before).} 
\label{zel1} 
\end{figure} 

\begin{figure} 
\centering 
\includegraphics[width=\hsize,clip]{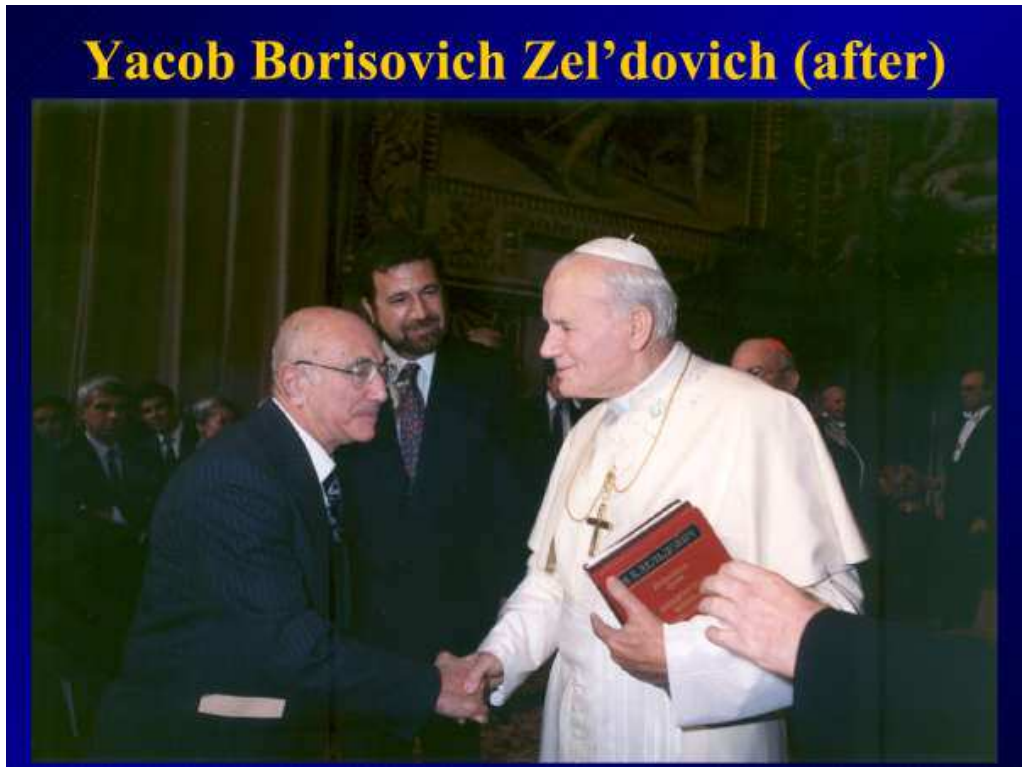} 
\caption{Y.B. Zel'dovich again shaking the hand of the Pope after presentation of his collected papers. Everybody smiles with relief!} 
\label{zel2} 
\end{figure}

I had my first meeting with Zel'dovich in 1968 in Tibilisi, Georgia in the former Soviet Union.
I was very impressed by his extensive scientific knowledge and at the same time intrigued by some peculiarities of his character, which immediately surfaced from the first scientific exchange and some anecdotes of his life that he recalled to me. Over the years we became very well acquainted and a great friendship developed between us. Nevertheless, this strangeness and somewhat unexpected manifestation of his character accompanied us all the way to our last meeting. It was in Rome that I had the occasion to introduce him to Pope John Paul II. Even in that solemn occasion Zel'dovich did not fail to manifest this duality between scientific knowledge and unexpected action. While he was approaching in line, I was introducing other distinguished guests to the Pope, including Bruno Pontecorvo, Roald Sagdeev and Rashid Sunyaev. I noticed that Zel'dovich had hidden under his jacket a voluminous object. This became more and more evident as he was approaching the Pope. You can see the concern in my eyes in Fig.~\ref{zel1}. When he arrived in front of the Pope, he suddenly opened the jacket and extracted two big red volumes and then handed them to the Pope. The Pope kindly thanked him. But again unexpectedly Zel'dovich took the volumes back from the hand of the Pope, and shouted loudly: ``Not just thank you, these are fifty years of my work!'' 
We all realized that these were his collected papers. We then all felt much more relaxed and warmly laughed with great relief. The Pope kept the Zel'dovich books under his arm against his white robe during the entire rest of the audience (see Fig. \ref{zel2}).

The topic I will speak about today is again related to the dual activity of Zel'dovich, and it is definitely one of the most astonishing proposals ever made by a {\it homo sapiens}, which led to the discovery of gamma ray bursts. It was during the 1950s that Zel'dovich, in order to show the greatness of his scientific achievements and the very large progress in space technology made by the Soviet Union, made a proposal to explode an atomic bomb on the Moon. In his opinion this would have shown the superiority of the Soviet rocketry to reach the Moon before the Americans and would have allowed a large fraction of inhabitants of the Earth to directly see this achievement by the observation of the fireball of the bomb explosion at a very precisely predicted time. 
Many technicalities hampered the realization of this idea: fortunately, this unacceptable proposal was never implemented. But the possibility of conceiving of such an action had become a reality, and the United States put the {\it Vela} satellites into very high Earth orbits in order to monitor the nonproliferation agreement. They discovered the gamma ray bursts. In this case, therefore, even this extravagant and nonscientific proposal of Zel'dovich finally did materialize, fortunately, in a great scientific discovery.

\section{The energetics of gamma ray bursts} 

GRBs were detected and studied for the first time 
using those {\em Vela} satellites, developed for military research to 
monitor the violations of the Limited Test Ban Treaty signed in 1963 
(see e.g. Strong \cite{s75}). It was clear from the early data of these 
satellites, which were put at $150,000$ miles from the surface of the Earth, 
that the GRBs did not originate either on the Earth or in the Solar System. 

\begin{figure} 
\centering 
\includegraphics[height=\hsize,clip,angle=90]{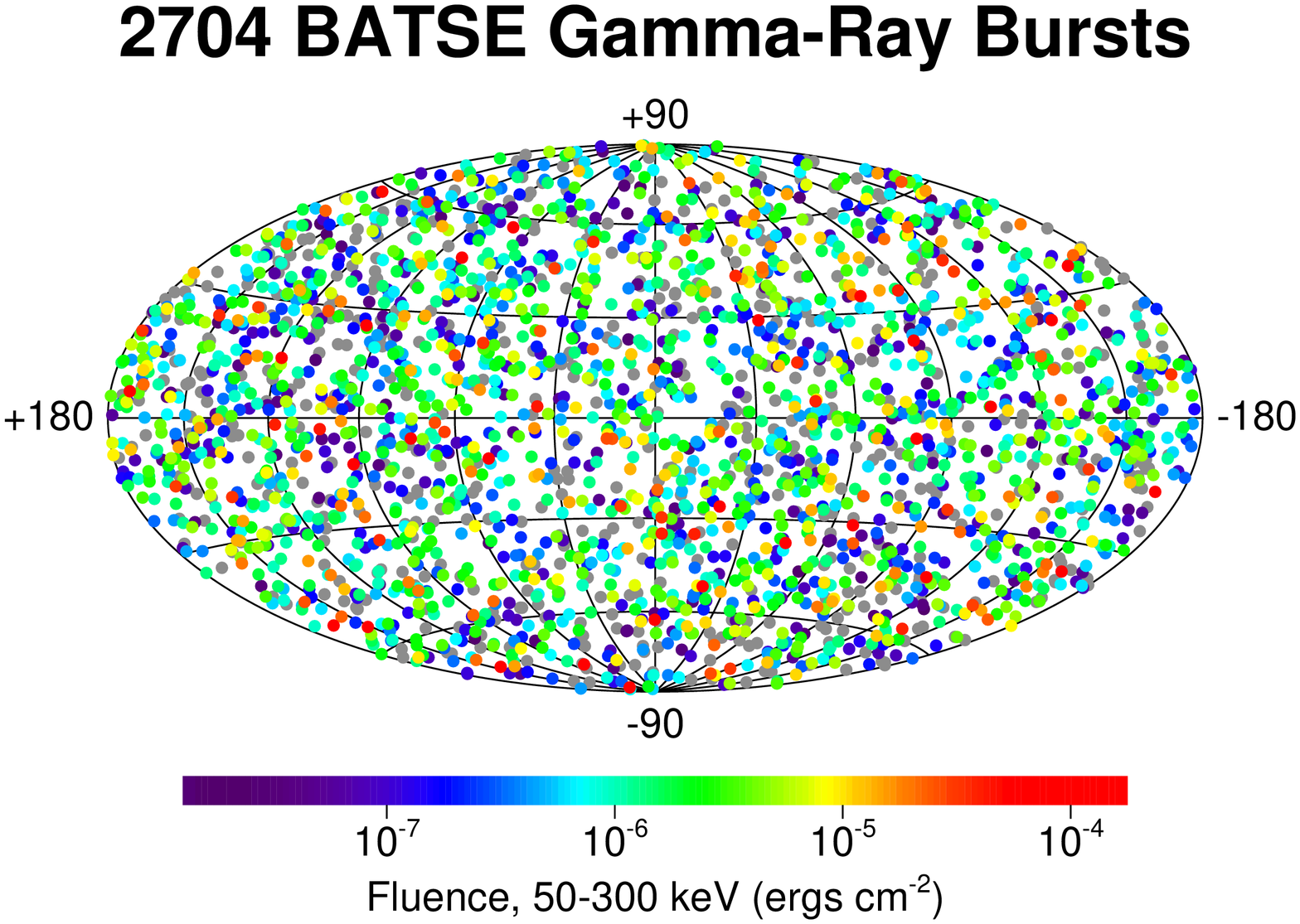} 
\caption{Position in the sky, in galactic coordinates, of 2000 GRB events 
seen by the CGRO satellite. Their isotropy is evident. Reproduced courtesy of the 
BATSE web site.} 
\label{batse2k} 
\end{figure} 

\begin{figure} 
\centering 
\includegraphics[width=\hsize,clip]{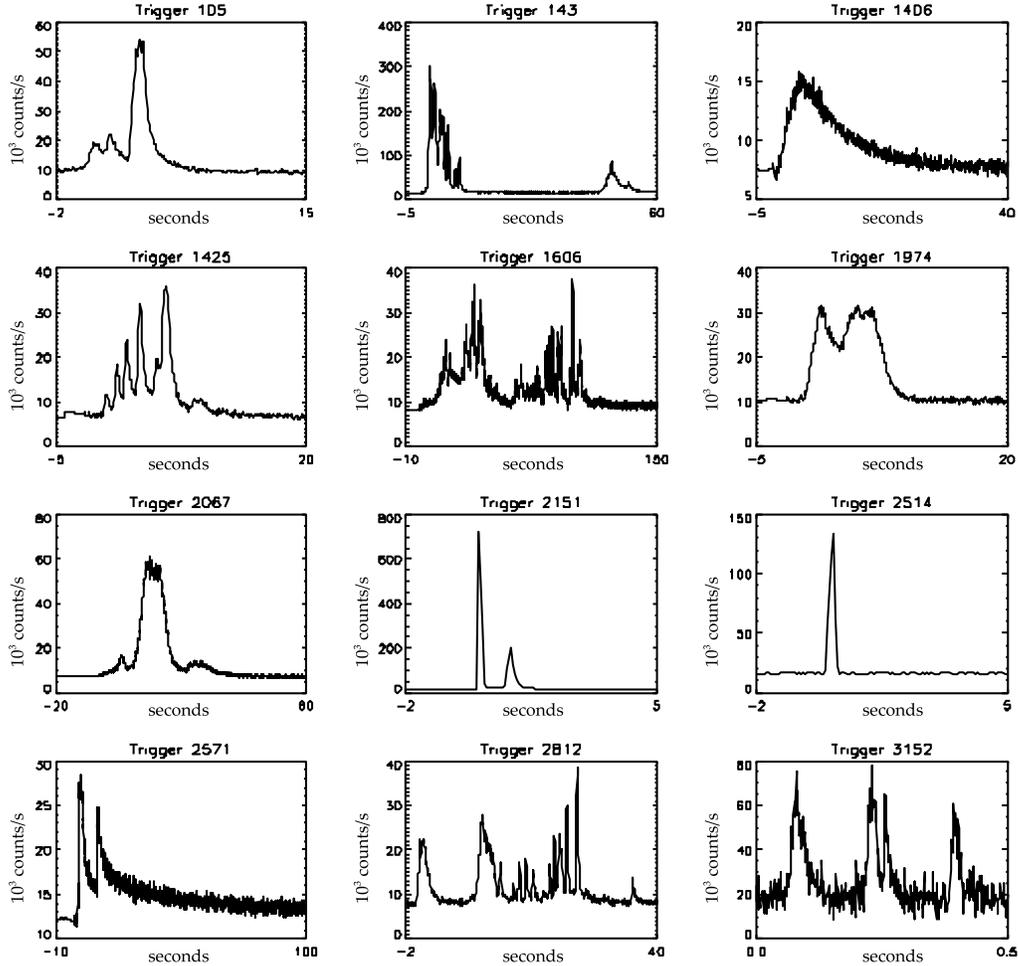} 
\caption{Some GRB light curves observed by the BATSE instrument on board 
the CGRO satellite.} 
\label{grb_profiles_eng} 
\end{figure}

\begin{figure} 
\centering 
\includegraphics[width=\hsize,clip]{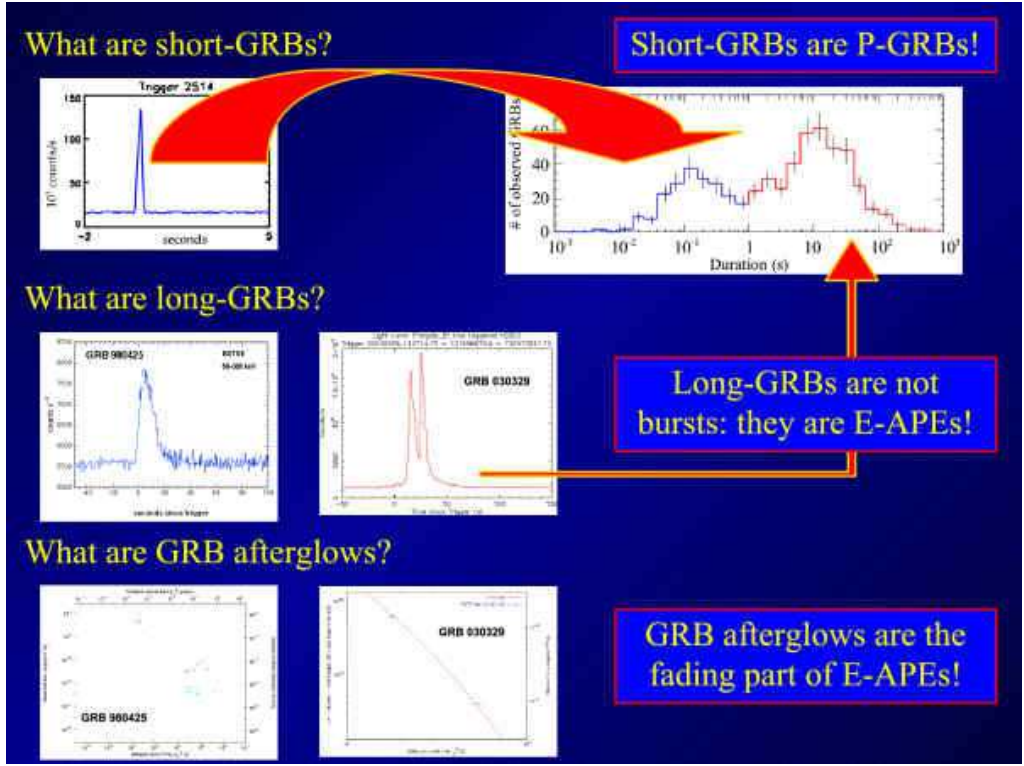} 
\caption{On the upper right part of the figure are plotted the number of 
the observed GRBs as a function of their duration. The bimodal 
distribution corresponding respectively to the short bursts, upper left 
figure, and the long bursts, middle figure, is quite evident. 
} 
\label{slb} 
\end{figure}

The mystery of these sources became more profound as the observations of 
the BATSE instrument on board the Compton Gamma Ray Observatory (CGRO) 
satellite \footnote{see http://cossc.gsfc.nasa.gov/batse/} over 9 years 
proved the isotropic distribution of these sources in the sky (See Fig.~\ref{batse2k}). 
In addition to these data, the CGRO satellite gave an unprecedented number 
of details about the structure of GRBs, and on their spectral properties and time 
variabilities which were recorded in the fourth BATSE catalog 
\cite{batse4b} (see e.g.\ Fig.~\ref{grb_profiles_eng}). Out of the analysis 
of these BATSE sources the existence of two distinct 
families of sources soon became clear (see e.g.\ Koveliotou et al.\ 
\cite{ka93}, Tavani et al.\ \cite{t98}): the long bursts, lasting more then one second and 
softer in spectra, and the short bursts, harder in spectra (see Fig.~\ref{slb}).  

\begin{figure} 
\centering 
\includegraphics[width=\hsize,clip]{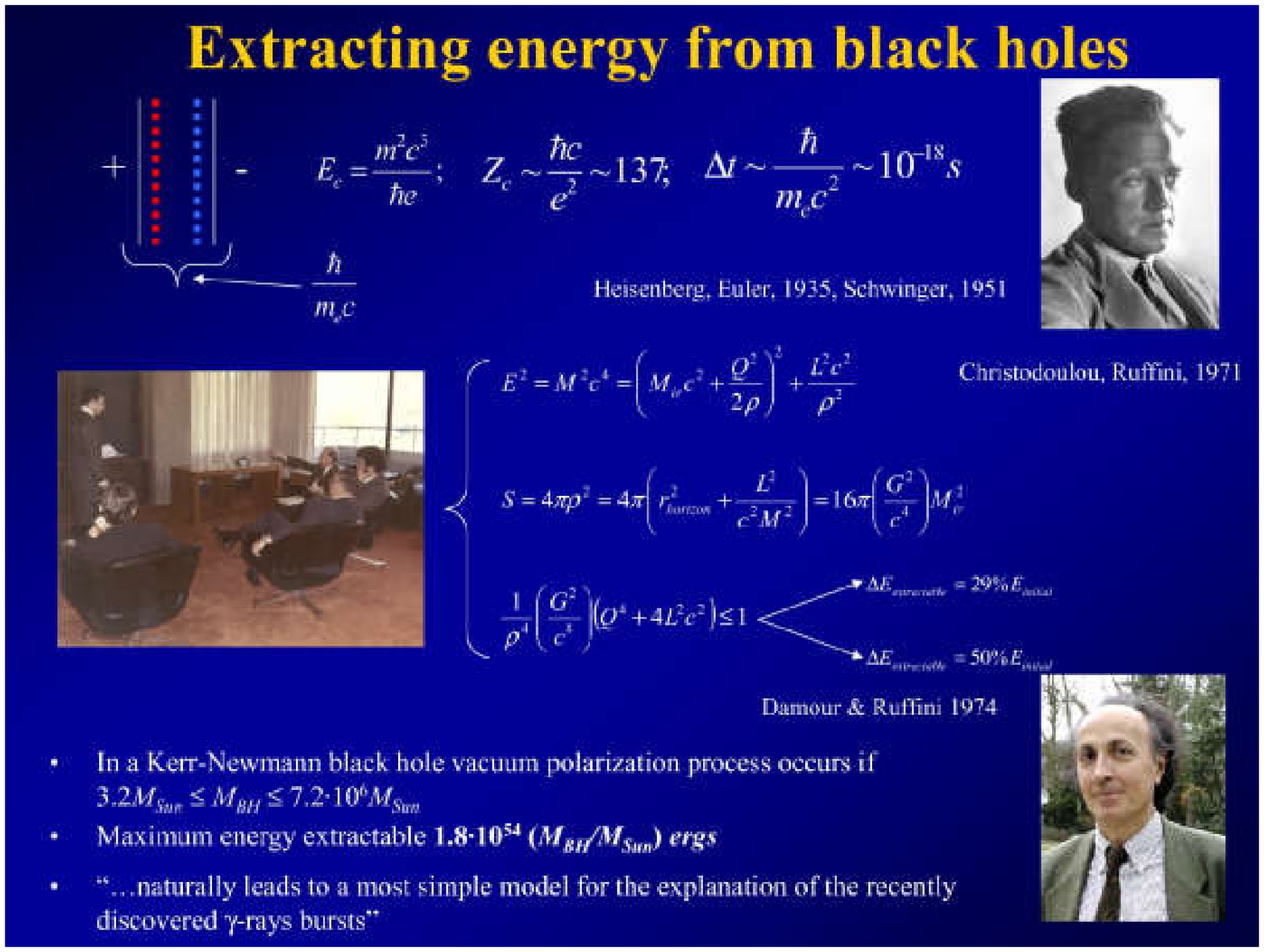} 
\caption{Damour} 
\label{damour}
\end{figure}

The situation drastically changed with the discovery of the afterglow by 
the Italian-Dutch satellite BeppoSAX (Costa et al. \cite{ca97}) and the 
possibility which led to the optical identification of the GRBs by the 
largest telescopes in the world, including the Hubble Space Telescope, the 
Keck Telescope in Hawaii and the VLT in Chile, and allowed as well the 
identification in the radio band of these sources. The outcome of this 
collaboration between complementary observational techniques has led in 1997  
to the possibility of identifying the distance of these sources from the 
Earth and their tremendous energy of the order up to $10^{54}$ erg/second 
during the burst. It is interesting, as we will show in the following, 
that energetics of this magnitude for the GRBs had  already been 
predicted out of first principles by Damour and Ruffini in 1974 
\cite{dr75} (see Fig. \ref{damour}).

The resonance between the X- and gamma ray astronomy from the satellites 
and the optical and radio astronomy from the ground, had already marked 
the great success and development of the astrophysics of 
binary X-ray sources in 
the seventies (see e.g. Giacconi \& Ruffini \cite{gr78}). This 
resonance has been repeated for GRBs on a much larger scale. The use of 
much larger satellites, like Chandra and XMM-Newton, and dedicated space 
missions, like HETE-2 and, in the near future, Swift, and the very 
fortunate circumstance of the coming of age of the development of 
unprecedented optical technologies for the telescopes offer opportunities 
without precedent in the history of mankind. In parallel, the enormous 
scientific interest in the nature of GRB sources and the exploration, not 
only of new regimes, but also of the totally novel conceptual physical process 
of blackholic energy extraction, makes the knowledge of GRBs an authentic new 
frontier in scientific knowledge. 

\subsection{GRBs and general relativity}\label{genrel} 

\begin{figure} 
\centering 
\includegraphics[width=\hsize,clip]{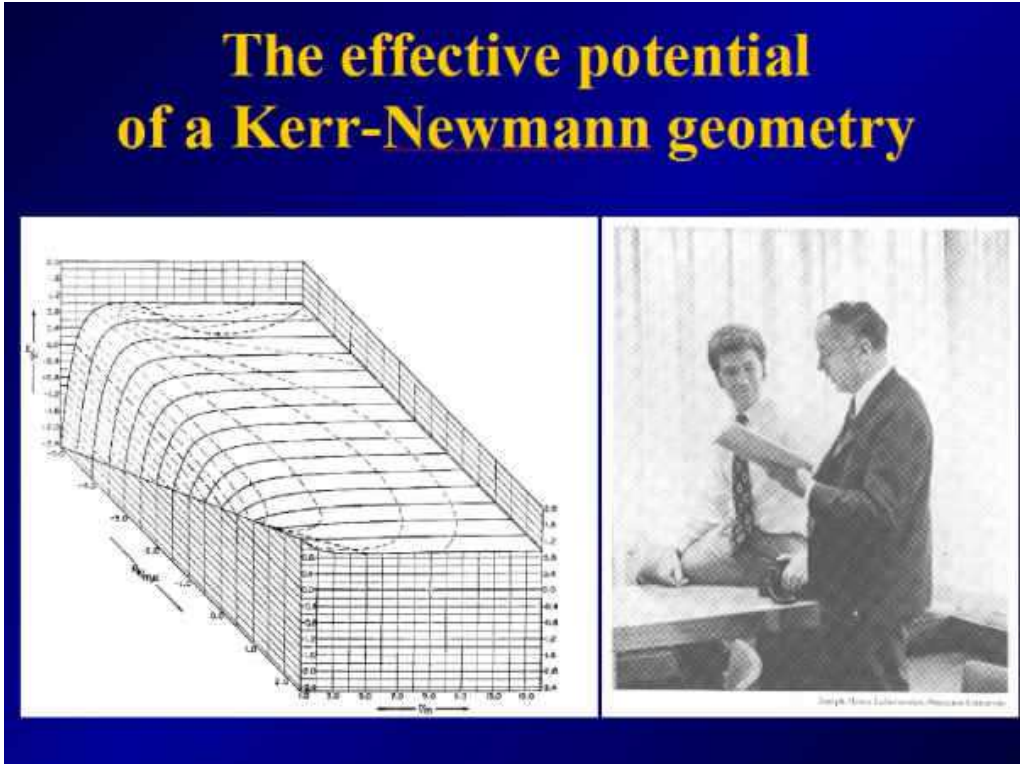} 
\caption{The effective potential corresponding to the circular orbits in 
the equatorial plane of a black hole is given as a function of the angular 
momentum of the test particle. This diagram was originally derived by 
Ruffini and Wheeler (right picture). For details see Landau and Lifshitz 
\cite{ll2} and Rees, Ruffini and Wheeler \cite{rrw}.} 
\label{effp} 
\end{figure}

Three of the most important works in the field of general relativity have 
certainly been the discovery of the Kerr solution \cite{kerr}, its 
generalization to the charged case (Newman et al.\ \cite{newman}) and the 
formulation by Brandon Carter \cite{carter} of the Hamilton-Jacobi 
equations for a charged test particle in the metric and electromagnetic 
field of a Kerr-Newman solution (see e.g.\ Landau and Lifshitz \cite{ll2}). 
The equations of motion, which are generally second order differential 
equations, were reduced by Carter to a set of first order differential 
equations which were then integrated using an effective potential 
technique by Ruffini and Wheeler for the Kerr metric (see e.g. Landau and 
Lifshitz \cite{ll2}) and by Ruffini for the Reissner-Nordstr\"om geometry 
(Ruffini \cite{r70}, see Fig. \ref{effp}). 

All the above mathematical results were essential for understanding the 
new physics of gravitationally collapsed objects and allowed the 
publication of a very popular article: ``Introducing the black hole'' 
(Ruffini and Wheeler \cite{rw71}). In that paper, we advanced the ansatz 
that the most general black hole is a solution of the Einstein-Maxwell 
equations, asymptotically flat and with a regular horizon: the Kerr-Newman 
solution, characterized only by three parameters: the mass $M$, the charge 
$Q$ and the angular momentum $L$. This ansatz of the ``black hole 
uniqueness theorem'' still today after thirty years presents challenges to 
the mathematical aspects of its complete proof (see e.g.\ Carter \cite{ckf} 
and Bini et al.\ \cite{bcjr}). In addition to these mathematical 
difficulties, in the field of physics this ansatz contains the most 
profound consequences. The fact that, among all the possible highly 
nonlinear terms characterizing the gravitationally collapsed objects, only 
the ones corresponding solely to the Einstein-Maxwell equations survive 
the formation of the horizon has, indeed, extremely profound physical 
implications. Any departure from such a minimal configuration either 
collapses to the horizon or is radiated away during the collapse process. 
This ansatz is crucial in identifying precisely the process of 
gravitational collapse leading to the formation of the black hole and the 
emission of GRBs. Indeed, in this specific case, the Born-like nonlinear term 
\cite{b33} of the Heisenberg-Euler-Schwinger Lagrangian \cite{he35,s51} 
are radiated away prior to the formation of the horizon of the 
black hole (see e.g.\ Ruffini et al.\ \cite{rvx05}). Only the nonlinearity 
corresponding solely to the classical Einstein-Maxwell theory is left as 
the outcome of the gravitational collapse process. 

The same effective potential technique (see Landau and Lifshitz 
\cite{ll2}) which allowed the analysis of circular orbits around the 
black hole was crucial in reaching the equally interesting discovery of 
the reversible and irreversible transformations of black holes by 
Christodoulou and Ruffini \cite{cr71}, which in turn led to the 
mass-energy formula for the black hole 
\begin{equation} 
E_{BH}^2 = M^2c^4 = \left(M_{\rm ir}c^2 
+ \frac{Q^2}{2\rho_+}\right)^2+\frac{L^2c^2}{\rho_+^2}\, , 
\label{em} 
\end{equation} 
with 
\begin{equation} 
\frac{1}{\rho_+^4}\left(\frac{G^2}{c^8}\right)\left(Q^4+4L^2c^2\right)\leq 
1\, , 
\label{s1} 
\end{equation} 
and where 
\begin{equation} 
S=4\pi\rho_+^2=4\pi(r_+^2+\frac{L^2}{c^2M^2})=16\pi\left(\frac{G^2}{c^4}\right) 
M^2_{\rm ir}\, , 
\label{sa} 
\end{equation} 
is the horizon surface area, $M_{\rm ir}$ is the irreducible mass, $r_{+}$ 
is the horizon radius and $\rho_+$ is the quasi-spheroidal cylindrical 
coordinate of the horizon evaluated at the equatorial plane. Extreme EMBHs 
satisfy the equality in Eq.~(\ref{s1}). 

From Eq.~(\ref{em}) follows that the total energy of the black hole 
$E_{BH}$ can be split into three different parts: rest mass, Coulomb 
energy and rotational energy. In principle both Coulomb energy and 
rotational energy can be extracted from the black hole (Christodoulou and 
Ruffini \cite{cr71}). The maximum extractable rotational energy is 29\% 
of the total energy and the maximum extractable Coulomb energy is 50\%, as 
clearly follows from the upper limit for the existence of a black hole, 
given by Eq.~(\ref{s1}). We refer to both these 
extractable energies in the following as the blackholic energy. 

The existence of the black hole and the basic correctness of the circular 
orbits has been proven by the observations of Cygnus-X1 (see e.g.\ Giacconi 
and Ruffini \cite{gr78}). However, in binary X-ray sources, the black hole 
only acts passively by generating the deep potential well in which the 
accretion process occurs. It has become tantalizing to look for 
astrophysical objects in order to verify the other fundamental prediction 
of general relativity that the blackholic energy is the largest energy 
extractable from any physical object. 

As we shall see in the next section, the feasibility of extracting 
the blackholic energy has been made possible by the quantum process of 
creating, out of classical fields, a plasma of electron-positron pairs in 
the field of a black hole. This process of energy 
extraction from the black hole is manifested astrophysically by the 
occurrence of GRBs. 

\subsection{GRBs and quantum electrodynamics} 

That a static electromagnetic field stronger than a critical value 
\begin{equation} 
E_c = \frac{m_e^2c^3}{\hbar e} 
\label{ec} 
\end{equation} 
can polarize the vacuum and create electron-positron pairs was clearly 
shown by Heisenberg and Euler \cite{he35}. The major effort in 
verifying the correctness of this theoretical prediction has been directed 
towards the analysis of heavy ion collisions (see Ruffini et al.\ \cite{rvx05} 
and references therein). From an order-of-magnitude estimate, it appears 
that around a nucleus with a charge 
\begin{equation} 
Z_c \simeq \frac{\hbar c}{e^2} \simeq 137 \, ,
\label{zc} 
\end{equation} 
the electric field can be stronger than the critical electric field 
needed to polarize the 
vacuum. A more accurate detailed analysis taking into account the bound 
state levels around a nucleus increases the value to 
\begin{equation} 
Z_c \simeq 173 
\label{zc2} 
\end{equation} 
for the nuclear charge leading to the existence of a critical field. From 
the Heisenberg uncertainty principle it follows that, in order to create a 
pair, the existence of the critical field should last a time 
\begin{equation} 
\Delta t \sim \frac{\hbar}{m_e c^2} \simeq 10^{-18}\, \mathrm{s}\, , 
\label{dt} 
\end{equation} 
which is much longer than the typical confinement time in heavy ion 
collisions which is 
\begin{equation} 
\Delta t \sim \frac{\hbar}{m_p c^2} \simeq 10^{-21}\, \mathrm{s}\, . 
\label{dt2} 
\end{equation} 
This is certainly a reason why no evidence for pair creation in heavy ion 
collisions has been found although remarkable efforts have been made in 
various accelerators worldwide. Similar experiments involving laser beams 
encounter analogous difficulties (see e.g. Ruffini et al. \cite{rvx05} and 
references therein). 

The alternative idea was advanced in 1975 \cite{dr75} that the critical 
field condition given in Eq.~(\ref{ec}) could be reached easily, and for a 
time much larger than the one given by Eq.~(\ref{dt}), in the field of a 
Kerr-Newman black hole in a range of masses $3.2M_\odot \le M_{BH} \le 
7.2\times 10^6M_\odot$. In that paper we generalized the fundamental 
theoretical framework developed in Minkowski space by Heisenberg-Euler 
\cite{he35} and Schwinger \cite{s51} to the curved 
Kerr-Newman geometry. 
This result was made possible by the work on the structure of the 
Kerr-Newman spacetime previously done by Carter \cite{carter} and by the 
remarkable mathematical craftsmanship of Thibault Damour then working with 
me as a post-doc in Princeton. 

The maximum energy extractable in such a process of creating a vast amount 
of electron-positron pairs around a black hole is given by 
\begin{equation} 
E_{max} = 1.8\times 10^{54} \left(M_{BH}/M_\odot\right)\, \mathrm{erg}\, 
\mathrm{.} 
\label{emax} 
\end{equation} 
We concluded in that paper that such a process ``naturally leads to a most 
simple model for the explanation of the recently discovered gamma ray 
bursts''. 

At that time, GRBs had not yet been optically identified and nothing was 
known about their distance and consequently about their energetics. 
Literally thousands of theories existed in order to explain them and it 
was impossible to establish a rational dialogue with such an enormous 
number of alternative theories. We did not pursue further our model until 
the results of the BeppoSAX mission, which clearly pointed to the 
cosmological origin of GRBs, implying for the typical magnitude of their 
energy precisely the one predicted by our model. 

It is interesting that the idea of using an electron-positron plasma as the 
basis of a GRB model was independently introduced years later in a set of 
papers by Cavallo and Rees \cite{cr78}, Cavallo and Horstman \cite{ch81} 
and Horstman and Cavallo \cite{hc83}. These authors did not address the 
issue of the physical origin of their energy source. They reach their 
conclusions considering the pair creation and annihilation process 
occurring in the confinement of a large amount of energy in a region of 
dimension $\sim 10$ km typical of a neutron star. No relation to the 
physics of black holes nor to the energy extraction process from a black 
hole was envisaged in their interesting considerations, mainly directed to 
the study of the opacity and the consequent dynamics of such an 
electron-positron plasma. 

\begin{figure} 
\centering 
\includegraphics[width=\hsize,clip]{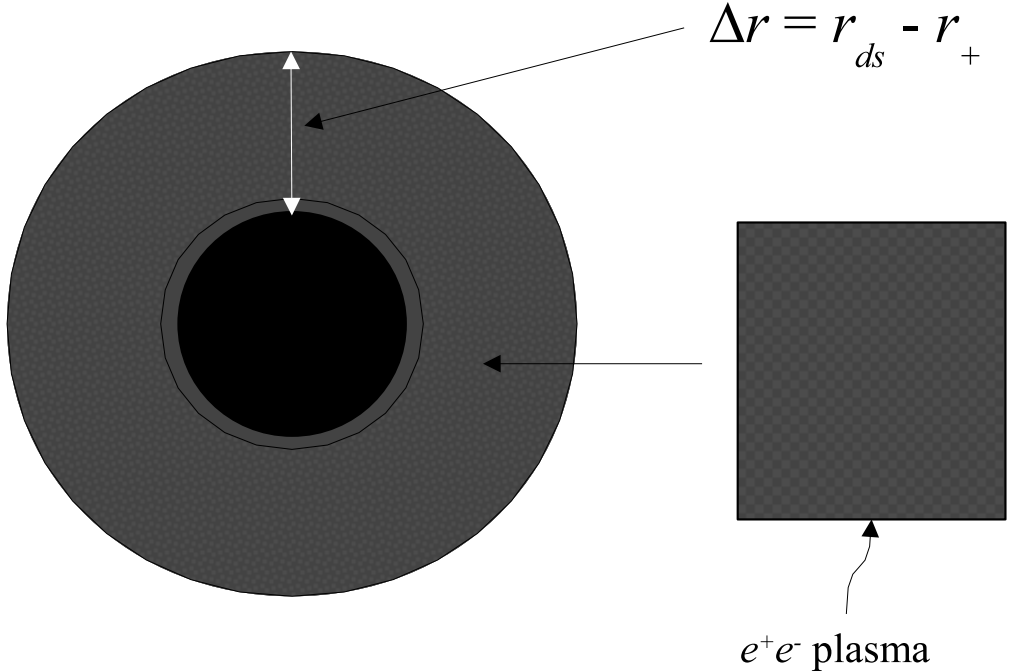} 
\caption{The dyadosphere is comprised between the horizon radius and the 
radius of the dyadosphere. This region is entirely filled with 
electron-positron pairs and photons in thermal equilibrium. Details in 
Ruffini \cite{rukyoto}, Preparata et al. \cite{prx98}, Ruffini et al. 
\cite{rvx03a}.} 
\label{dya} 
\end{figure} 

After the discovery of the afterglows and the optical identification of 
GRBs at cosmological distances, implying exactly the energetics predicted 
in Eq.~(\ref{emax}), we returned to the analysis of the vacuum polarization 
process around a black hole and precisely identified the region around the 
black hole in which the vacuum polarization process and the subsequent 
creation of electron-positron pairs occur. We defined this region, using 
the Greek name dyad for pairs ($\delta\upsilon\alpha\varsigma$, 
$\delta\upsilon\alpha\delta o \varsigma$), to be the ``dyadosphere'' of 
the black hole, bounded  by the black hole horizon and the dyadosphere 
radius $r_{ds}$ given by (see Ruffini \cite{rukyoto}, Preparata et al. 
\cite{prx98} and Fig.~\ref{dya}) 
\begin{equation} 
r_{ds}=\left(\frac{\hbar}{mc}\right)^\frac{1}{2}\left(\frac{GM}{ 
c^2}\right)^\frac{1}{2} \left(\frac{m_{\rm 
p}}{m}\right)^\frac{1}{2}\left(\frac{e}{q_{\rm 
p}}\right)^\frac{1}{2}\left(\frac{Q}{\sqrt{G}M}\right)^\frac{1}{2}
=1.12\cdot 
10^8\sqrt{\mu\xi} \, {\rm cm}\, , 
\label{rc} 
\end{equation} 
where we have introduced the dimensionless mass and charge parameters 
$\mu={M_{BH}/M_{\odot}}$, $\xi={Q/(M_{BH}\sqrt{G})}\le 1$. 

At that time the analysis of the dyadosphere was developed  around an 
already formed black hole. In recent months we have been developing the 
dynamical formation of the black hole and correspondingly of the 
dyadosphere during the process of gravitational collapse, reaching some 
specific signatures which may be detectable in the structure of the short 
and long GRBs (Cherubini et al. \cite{crv02}, Ruffini and Vitagliano 
\cite{rv02a,rv02b}, Ruffini et al. \cite{rvx03a,rvx03b,rfvx05}). 

\section{Reconsideration of a classic Fermi paper}

At the very foundation of the GRB phenomena is the vacuum polarization process due to overcritical electric fields of black holes. For these reasons we decided to go back to some of our earlier work and some other classic work in the literature on test particles in gravitational fields, and we have discovered a wealth of new results and opened as well additional new problems for enquiry. 
We have reconsidered a pioneering paper by Enrico Fermi \cite{fermi}, which has been generally neglected since it was written in Italian. It has only just now been translated into English \cite{fermibook}.
In this paper Fermi investigated the electric field generated by a charged particle at rest in a given static and homogeneous gravitational field in the spacetime region close to the particle location and then used his result to study the 
influence of the gravitational field on the charge distribution on
an infinitely conducting sphere. He showed that in this case the sphere acquires a dipole electric field and is polarized. In fact, the solution
for the electrostatic potential (and the field) can be expressed as the superposition of the solutions corresponding to a point charge and a dipole of suitable moment, in order to satisfy the condition of constancy of the potential on the surface of the sphere.

Fermi uses in his paper the following form of the metric due to Levi-Civita \cite{levicivita}  
for a uniform gravitational field
$$
\begin{array}{lc}
\label{Fmetric}
(\rm F1)\hspace{1cm}&  \rmd s^2 =-(1-2AZ)\rmd T^2 + \rmd X^2+\rmd Y^2+\rmd Z^2 +O(AZ)\, , \\
\end{array}
$$
with the condition $AZ\ll 1$; $A$ denotes the acceleration of gravity.
In this metric Fermi considers Maxwell's equations
$$
\begin{array}{lc}
\label{maxwell}
(\rm F2)\hspace{1cm}&  F^{\alpha\beta}{}_{;\beta}=4\pi J^\alpha, \qquad {}^* F^{\alpha\beta}{}_{;\beta}=0\ , 
\qquad 
F_{\alpha\beta}=2A_{[\beta;\alpha]}\, .
\end{array}
$$
One can introduce (pseudo-) electric and magnetic field quantities 
$$
\begin{array}{lc}
(\rm F3)\hspace{4.5cm}&  E_i=F_{i0}\, , \qquad 
B^i=\frac12\epsilon^{ijk}F_{jk}\, .
\end{array}
$$
differing from the physical fields which are instead the orthonormal frame components (and not the coordinate components) of the Faraday 2-form $F$.
In the electrostatic case $B_i=0$ and $E_{i,0}=0$, and the vector potential $A_\mu$ is determined by the electrostatic potential $\Phi$ alone ($A_0=\Phi\ , A_i=0$), 
in terms of which the (pseudo-) electric field components can be written in the form $E_X=-\Phi_{,X}, E_Y=-\Phi_{,Y}, E_Z=-\Phi_{,Z}$.

He then considers a charge $q$ located at the origin of the coordinates as the source term $J^\alpha$, i.e., with current density
$$
\begin{array}{lc}
\label{jfermi}
(\rm F4)\hspace{3.3cm}&  J^\alpha=\rho u^\alpha\ , \qquad  \rho=q\delta(X)\delta(Y)\delta(Z)\, , \\
\end{array}
$$
where $u=(1-AZ)^{-1}\partial_T$ is the particle 4-velocity.

In the limit of validity of the metric (F1), Maxwell's equations reduce to the following equation for the electrostatic potential $\Phi$ 
$$
\begin{array}{lc}
\label{fermipoteq}
(\rm F5)\hspace{3.9cm}&  \nabla^2 \Phi+A\partial_Z \Phi=-4\pi (1-AZ)\rho\, .
\end{array}
$$
The solution corresponding to the source term (F4) is given by
$$
\begin{array}{lc}
\label{fermipart}
(\rm F6)\hspace{1.8cm}&  \Phi_{\rm part}=\displaystyle\frac{q}{\sqrt{X^2+Y^2+Z^2}}\left[1-\frac{A}{2}Z\right]\, .
\end{array}
$$
Finally, let us assume that the charge is distributed on a conducting sphere of radius $R$, centered at the origin of the coordinate system.
From the condition that the electrostatic potential must be constant on the surface, Fermi finds that a polarization charge density appears on the sphere corresponding to a superposition of a monopolar and a dipolar distribution, explicitly given by 
$$
\begin{array}{lc}
\label{sigmafermi}
(\rm F7)\hspace{3.2cm}&  \sigma_F^R(\theta)=\displaystyle\frac{q}{4\pi R^2}+ \frac{qA}{2\pi R}\cos\theta\, . 
\end{array}
$$

Fermi also shows that the electric part of the electromagnetic field generated by the electric charge at rest in a homogeneous field of 
strength $A$ is equal to the electric part of the electromagnetic field which the same charge would produce in the absence of a gravitational field
if it moved in accelerated motion with acceleration $A/2$ in the opposite direction with respect to the gravitational field. 
However, a nonzero magnetic field is present in this latter case, and the Fermi solution corresponds to choosing suitably a gauge in such a way that the vector potential $A_\mu$ has the component $A_0=\Phi$ only.   
It is interesting that even in this problem there are still some open questions: the factor $1/2$ appearing in the acceleration $A/2$ still remains to be completely understood and is very likely connected by some analogy to the same factor 2 occurring in the Thomas precession. This topic is still a matter of active discussion with V. Belinski, D. Bini, J. Elhers and A. Geralico.

\section{Discussions of Wheeler-Hanni-Unruh on the charge near a Schwarzschild black hole}

\begin{figure} 
\centering 
\includegraphics[width=\hsize,clip]{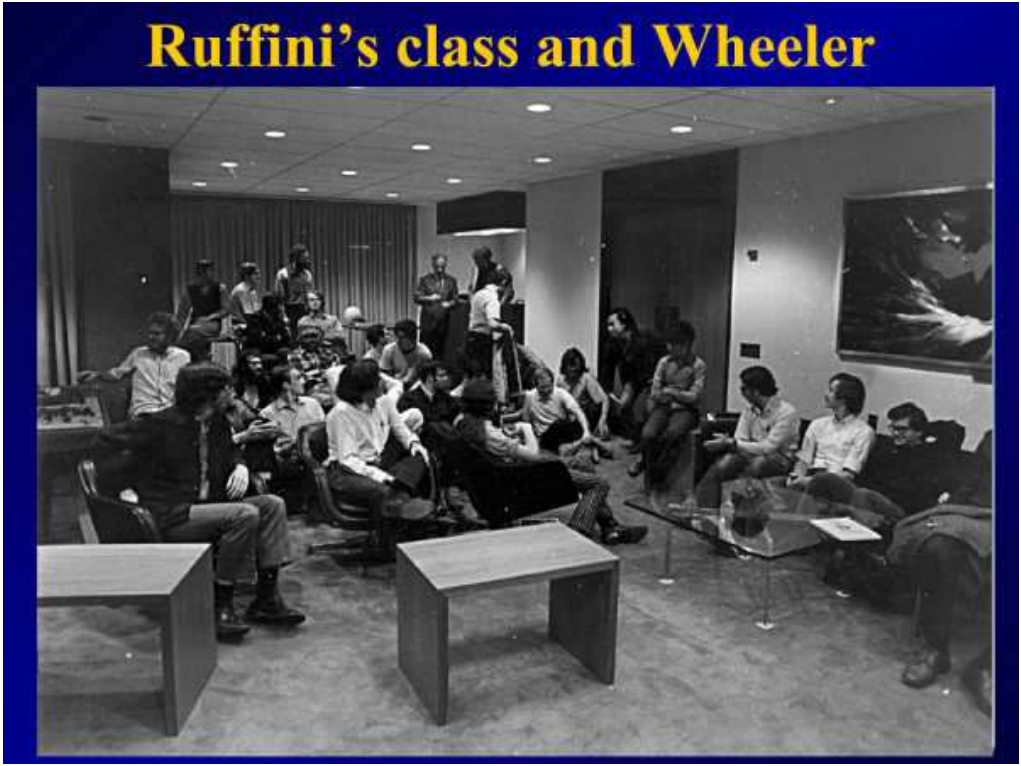} 
\caption{My students at Princeton in a evening discussion with John Wheeler. Recognizable on the right are Jim Eisenberg and Rick Hanni. Johnny and I are standing in the back of the room.} 
\label{class} 
\end{figure} 

\begin{figure} 
\typeout{*** EPS figure 1}
\begin{center}
$\begin{array}{cc}
\includegraphics[scale=0.30]{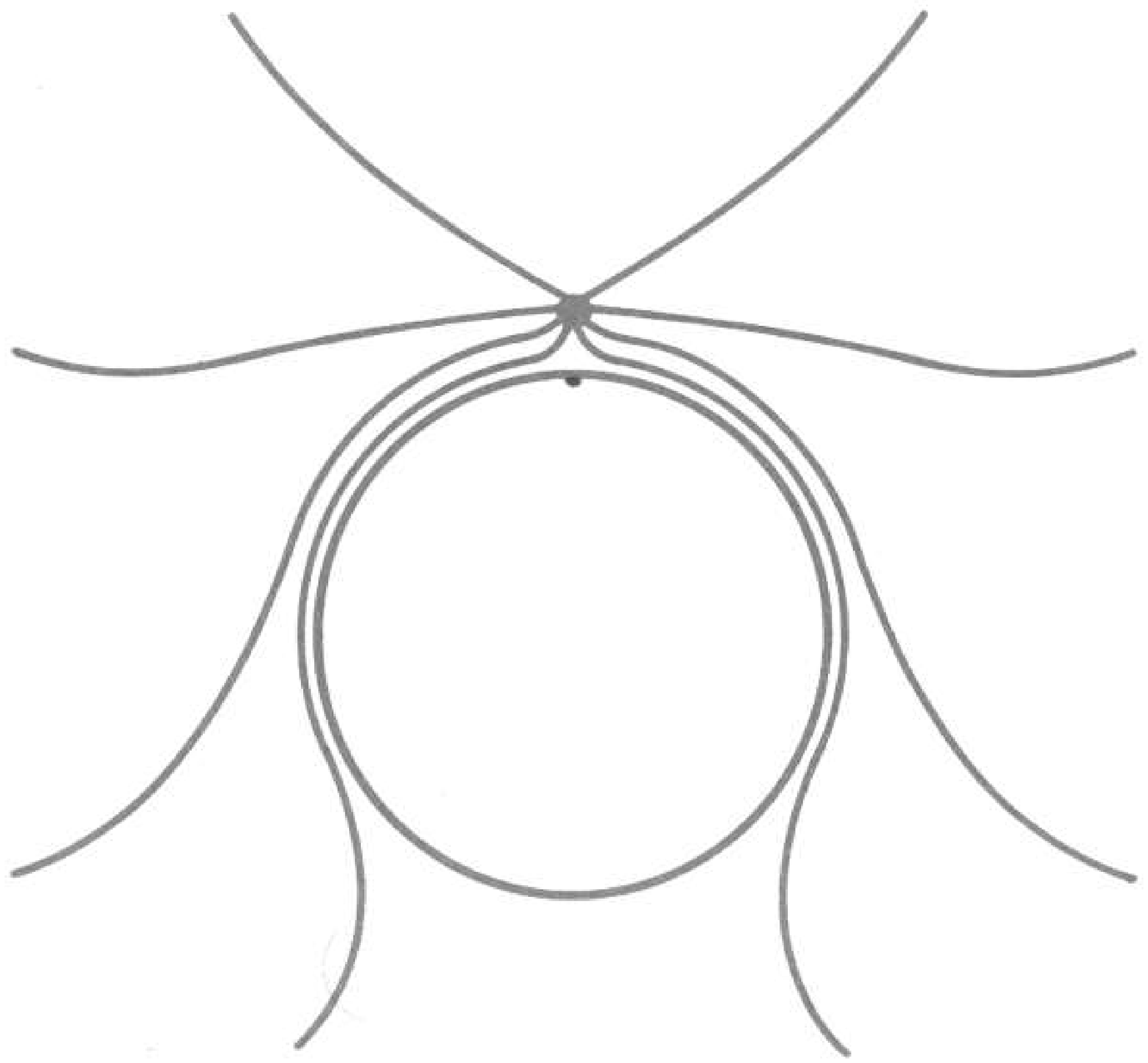}&
\includegraphics[scale=0.25]{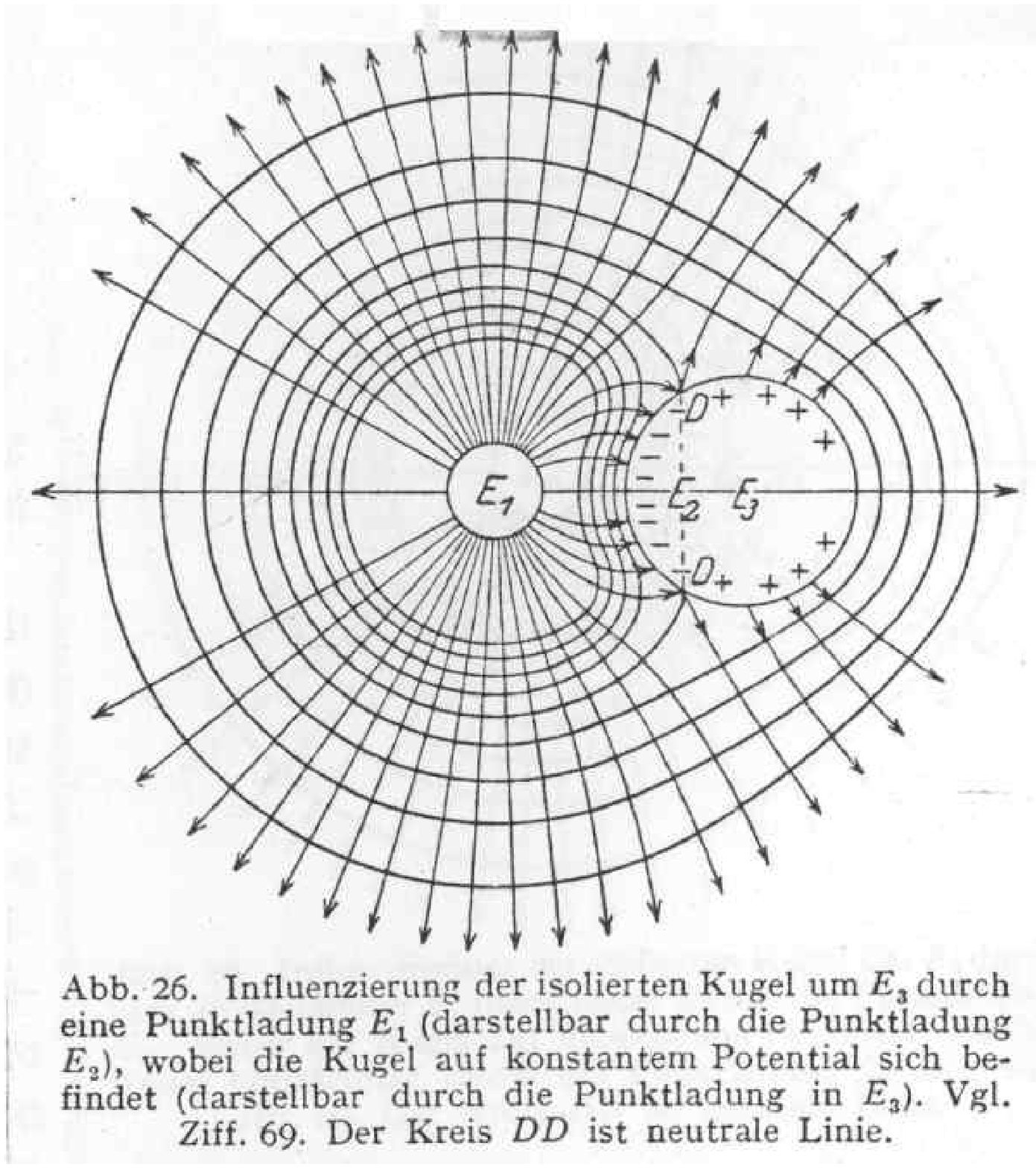}\\[0.4cm]
\mbox{(a)} & \mbox{(b)}\\
\end{array}$
\end{center}
\caption{The behavior of the lines of force of the electric field of the test particle as suggested by Wheeler and Unruh (Figs.~(a) and (b) respectively). The figure on the right is taken from the classical book of Weber on electricity and magnetism.}
\label{fig:1}
\end{figure}

One of the most exciting problems proposed by Johnny Wheeler to students and collaborators at Princeton (see Fig.~\ref{class}) was the problem of a charged test particle at rest near a black hole. The characteristic style of Wheeler has always been to have a strong intuition about the solution of physical problems. His motto, known as ``Wheeler theorem number 1'', is ``Never do a computation without knowing the solution,'' and he was usually extremely good at guessing the solution of a problem. 
We had just introduced with Johnny the astrophysical concept of a black hole \cite{rw71}. It is interesting that the specific case of the charge near a black hole really caught the attention of the students at Princeton, and all of them participated in trying to find solutions of this problem. In particular, Jacob Bekenstein, William Unruh and many others contributed to a lively discussion on the possible outcome of the solution. There were two different possibilities for the field line configurations, as outlined in Fig.~\ref{fig:1}: the one on the left was the first proposal of Johnny, and the one on the right was the one proposed by Unruh, adopting for the Schwarzschild black hole the analogy with an infinitely conducting metal sphere.


\begin{figure}
\typeout{*** EPS figure hanni}
\centering
\includegraphics[width=\hsize,clip]{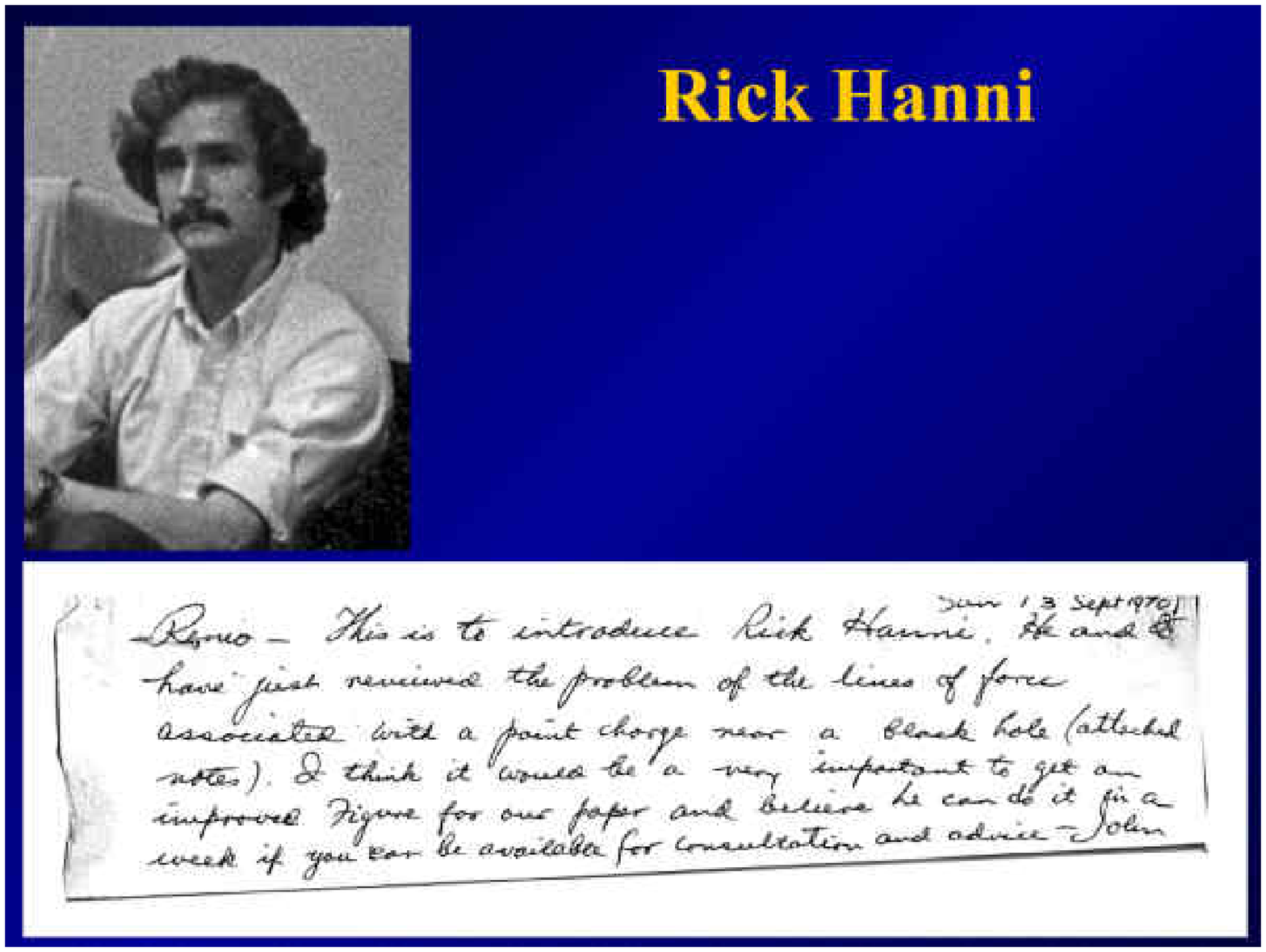}
\caption{Rick Hanni at Princeton and the note of Wheeler introducing him. Johnny optimistically, as usual, was expecting the problem of the lines of force to be solved in a week. It took almost one year of very hard work \cite{HR}.}
\label{fotohanni}
\end{figure}

While the discussions were polarizing our small scientific community in Princeton, I decided to enter in this issue by bypassing the philosophical and intuitive approach and just solve the corresponding set of equations. It was at that time that Wheeler introduced me to a young very bright undergraduate, Rickard Hanni (see Fig. \ref{fotohanni}).


\begin{figure}
\typeout{*** EPS figure hanni2}
\centering
\includegraphics[width=\hsize,clip]{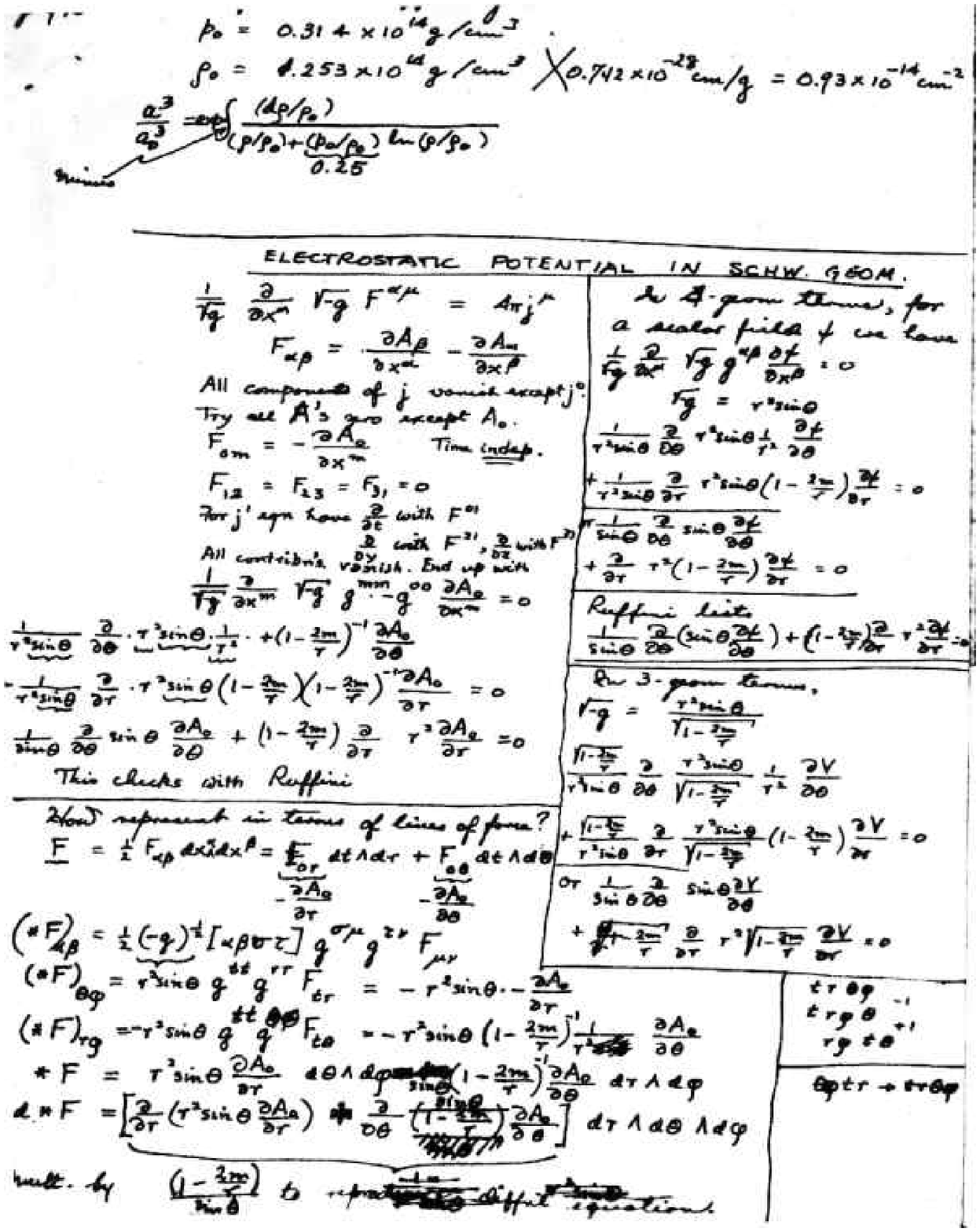}
\caption{From Wheeler's notebook (1).}
\label{eqshanni1}
\end{figure}


\begin{figure}
\typeout{*** EPS figure hanni2}
\centering
\includegraphics[width=\hsize,clip]{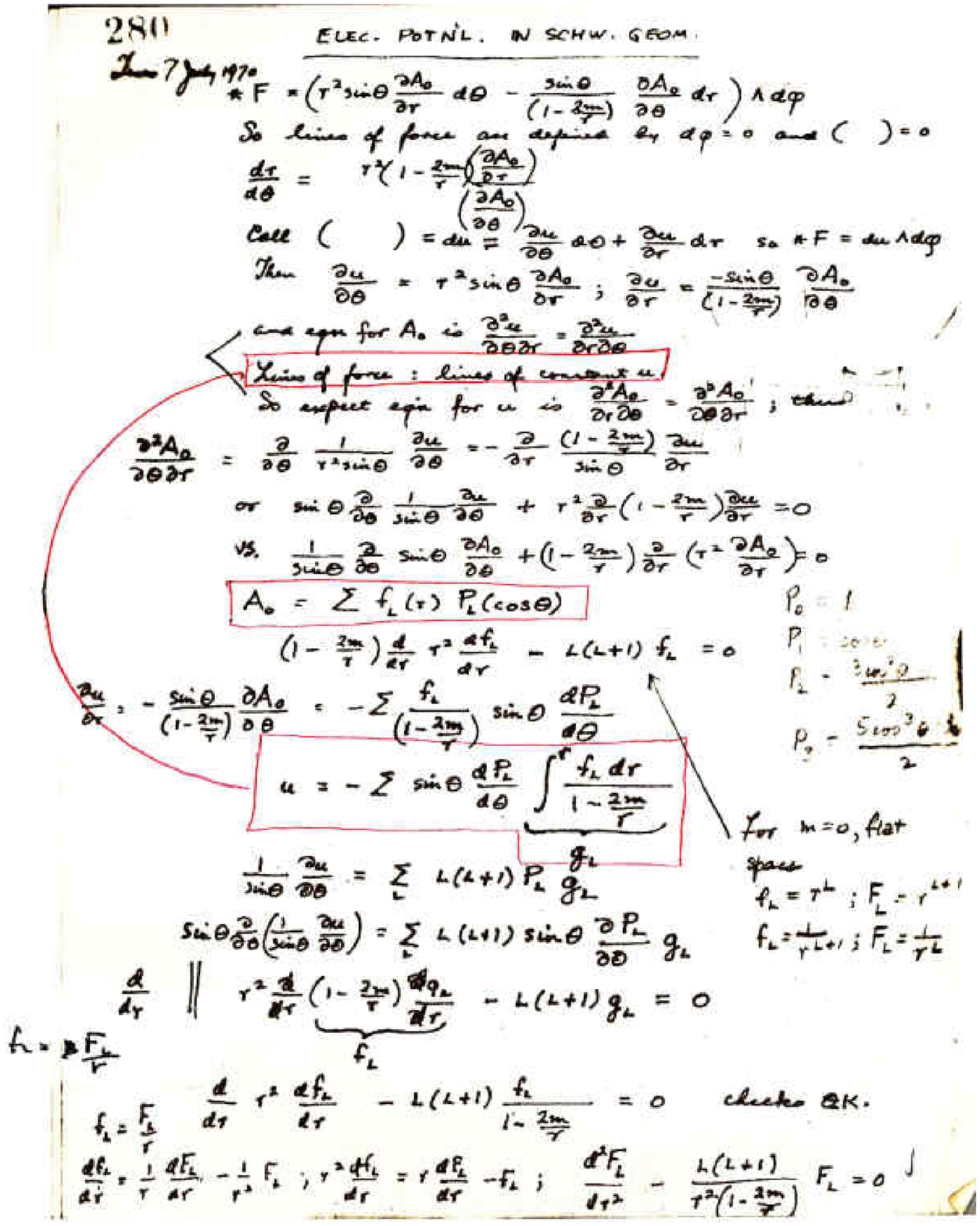}
\caption{From Wheeler's notebook (2).}
\label{eqshanni2}
\end{figure}

\begin{figure} 
\centering
\includegraphics[width=\hsize]{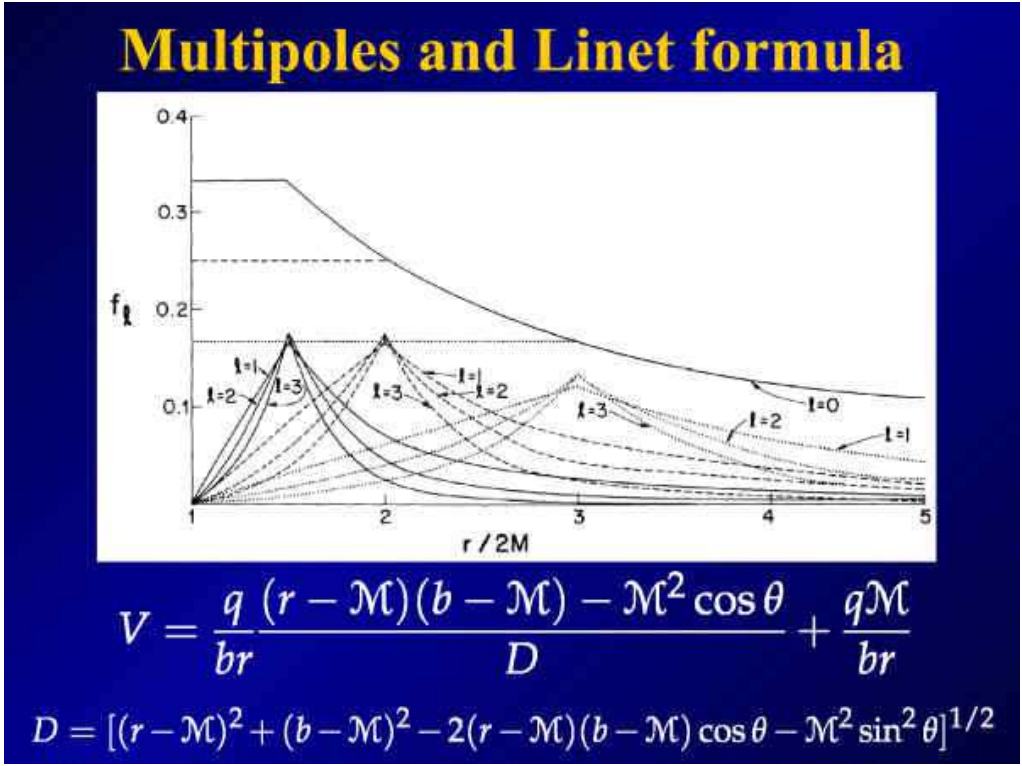}
\caption{The radial functions $f_l(r)$ in the multipole moment expression for the potential $V$ given by Eq.~(\ref{multipolischw}) are shown for a test particle at a selected distance from a Schwarzschild black hole. Below is the closed form of the electrostatic potential $V$ derived by Linet \cite{linet}.}
\label{solhanni}
\end{figure}

\begin{figure} 
\typeout{*** EPS figure 1}
\begin{center}
\includegraphics[scale=0.85]{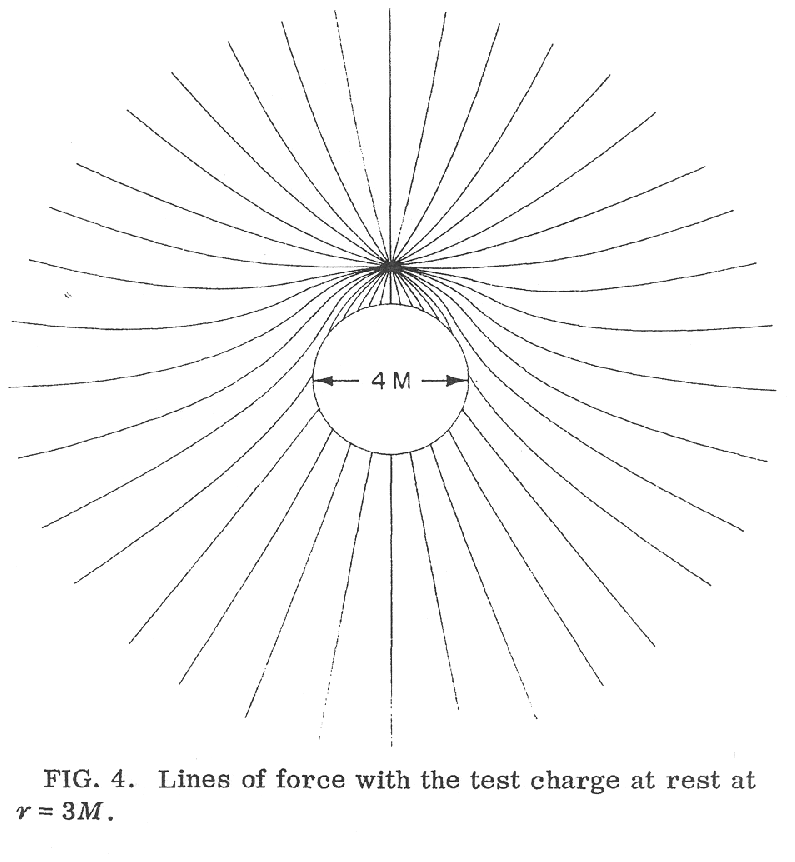}
\end{center}
\caption{Lines of force of the test field with the charged particle at rest on the vertical axis $\theta=0$ at $r=b $ with $b/{\mathcal M} = 3$ (from \cite{HR}). 
}
\label{fig:2}
\end{figure}

Recently, reading Fermi's paper, I noticed that the equations we used (see Figs.~\ref{eqshanni1} and \ref{eqshanni2}) have the same structure of the ones he used there, except instead of the Levi-Civita metric (F1), describing a uniform gravitational field, we used the Schwarzschild metric
$$
\begin{array}{lc}
\label{Smetric}
(\rm HR1)\hspace{1.5cm}&  ds^2=- f_{S}(r)dt^2 + f_{S}(r)^{-1}dr^2+r^2(d\theta^2 +\sin ^2\theta d\phi^2)\, ,
\end{array}
$$
where $f_{S}(r)=1-2\mathcal{M}/r$.
The current density corresponding to a charge $q$ placed 
at the point $r=b $ on the polar axis $\theta=0$, with $b > 2{\mathcal M}$, is given by
$$
\begin{array}{lc}
\label{Jzero}
(\rm HR4)\hspace{5cm}&  J^0=\displaystyle\frac q{2\pi r^2}\delta(r-b )\delta(\cos\theta-1)\, . 
\end{array}
$$
Maxwell's equations (F2) then reduce to the following equation for the electrostatic potential $V$ for the (pseudo-) electric field $E_r=-V_{,r}\ , E_\theta=-V_{,\theta}\ , E_\phi=-V_{,\phi}$, namely 
$$
\begin{array}{lc}
\label{poteqschw}
(\rm HR5)\hspace{.6cm}&  \displaystyle(r^2V_{,r})_{,r} +\frac{f_{S}(r)^{-1}}{\sin\theta}
\left[(\sin\theta V_{,\theta})_{,\theta} + \frac1{\sin\theta}(V_{,\phi})_{,\phi}\right]
=-4\pi r^2J^0\, .
\end{array}
$$
I solved this equation with Hanni (see Fig.~\ref{solhanni}) using a multipole expansion \cite{HR}
$$
\begin{array}{lc}
(\rm HR6)\hspace{1.1cm}&  V=q\sum_l [f_l(b)g_l(r)\vartheta(b-r)+g_l(b)f_l(r) \vartheta(r-b)]
P_l(\cos\theta)\, ,
\end{array}
$$
where
\begin{eqnarray}
\label{multipolischw}
f_l(r)&=&-\frac{(2l+1)!}{2^l(l+1)!l!{\mathcal M}^{(l+1)}}\frac{(r-2{\mathcal M})^2}r 
      \frac{dQ_l}{dr}\left[\frac{r-\mathcal M}{\mathcal M}\right]\qquad\,\,\,\, l=0,1,2, \ldots
\nonumber\\
g_l(r)&=&\left\{
     \begin{array}{ll}
     1&\qquad\quad l=0\\
     \noalign{\medskip}\displaystyle\frac{2^ll!(l-1)!{\mathcal M}^l}{(2l)!}\frac{(r-2{\mathcal M})^2}r 
     \frac{dP_l}{dr}\left[\frac{r-\mathcal M}{\mathcal M}\right]&\qquad\quad l=1,2, \ldots 
     \end{array}
     \right. \ 
\end{eqnarray}
and $P_l$, $Q_l$ are the Legendre functions. We then derived the lines of force by defining the lines of constant flux, obtaining the behavior shown in Fig.~\ref{fig:2} (details in \cite{HR}).

We also defined the concept of the induced charge on the surface of the black hole horizon, which indeed appears to have some of the properties of a perfectly conducting sphere. 
If we assume
that the test charge and black hole charge are both positive, at angles smaller than a certain critical angle the induced charge is negative and 
the lines of force go towards the horizon, while at angles greater than the critical angle the induced charge 
is positive and the lines of force go away from it. At the critical angle the induced charge density vanishes  
and the lines of force of the electric field are tangent to the horizon.

This confirmed the Unruh ansatz for the behaviour of the lines of force, but there always remained in my mind the question: ``How could it be that the very fertile imagination of Johnny did not enable him to guess a priori the correct solution?'' As we will show below recent developments may explain that at a deeper level Wheeler was indeed correct to be undecided on this issue. 
In the meantime my work with Hanni was improved by an important mathematical solution obtained by Linet \cite{linet}. He derived a closed form for the electrostatic potential (HR6) by summing over all the multipoles
\begin{equation}
\label{solschwpot}
V= \frac q{b r} \frac{(r-{\mathcal M})(b-{\mathcal M})
 -{\mathcal M}^2\cos\theta}{D_{S}} + \frac{q{\mathcal M}}{b r}\, ,
\end{equation}
with
\begin{equation}
\label{denschw}
D_{S}
= [(r-{\mathcal M})^2+(b-{\mathcal M})^2 
- 2(r -{\mathcal M})(b-{\mathcal M})\cos\theta 
- {\mathcal M}^2\sin^2\theta]^{1/2}\, . 
\end{equation}
The induced charge density on the horizon is then easily evaluated
$$
\begin{array}{lc}
(\rm HR7)\hspace{2.2cm}&  \sigma^H_{S}(\theta)
=\displaystyle\frac{q}{8\pi b} \frac{{\mathcal M}(1+\cos^2\theta)-2(b-\mathcal M)\cos\theta}
                     {[b-\mathcal M(1+\cos\theta)]^2}\, .
\end{array}
$$

\begin{figure} 
\typeout{*** EPS figure 2b}
\begin{center}
$\begin{array}{c@{\hspace{1in}}c}
\includegraphics[scale=0.5]{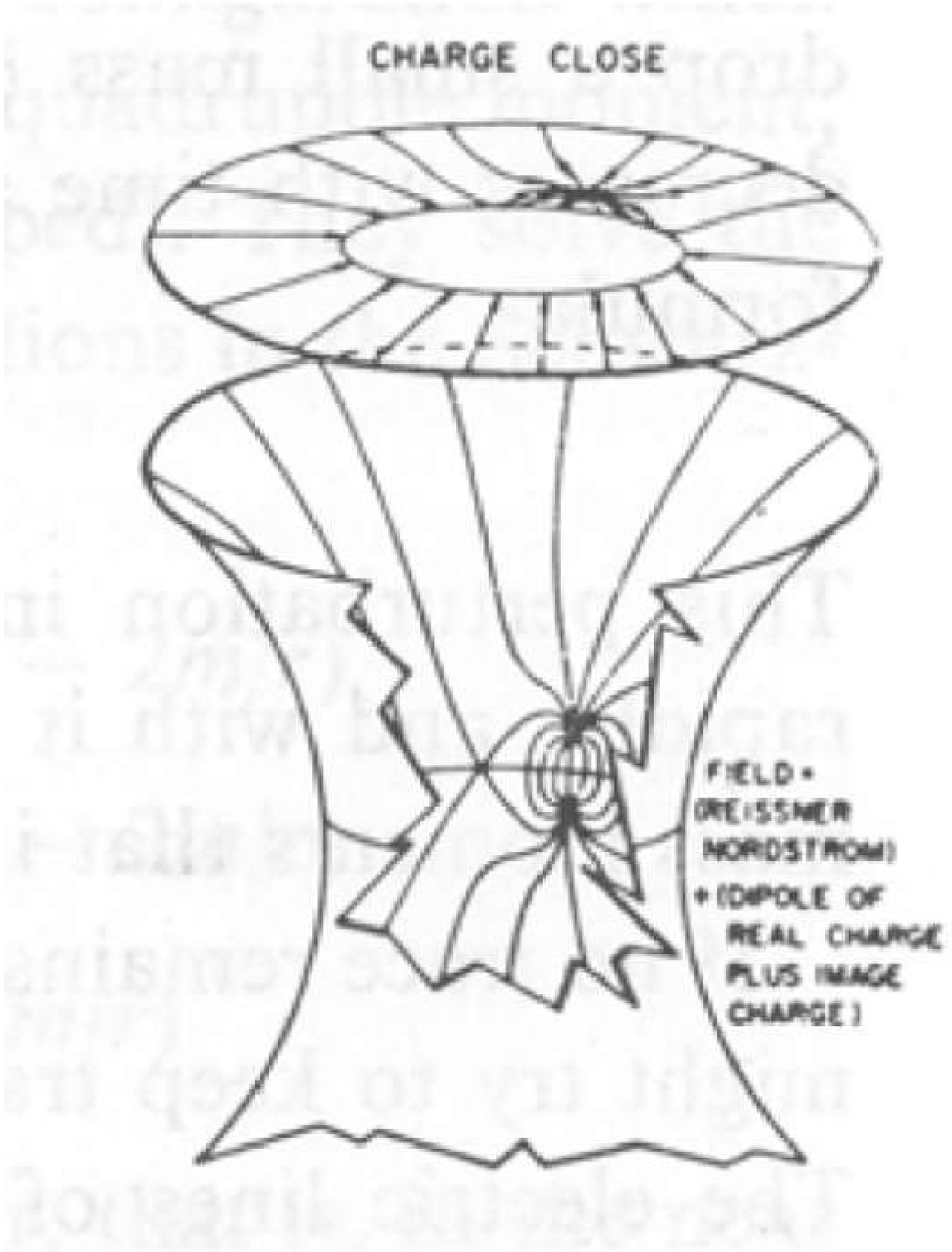}&
\includegraphics[scale=0.4]{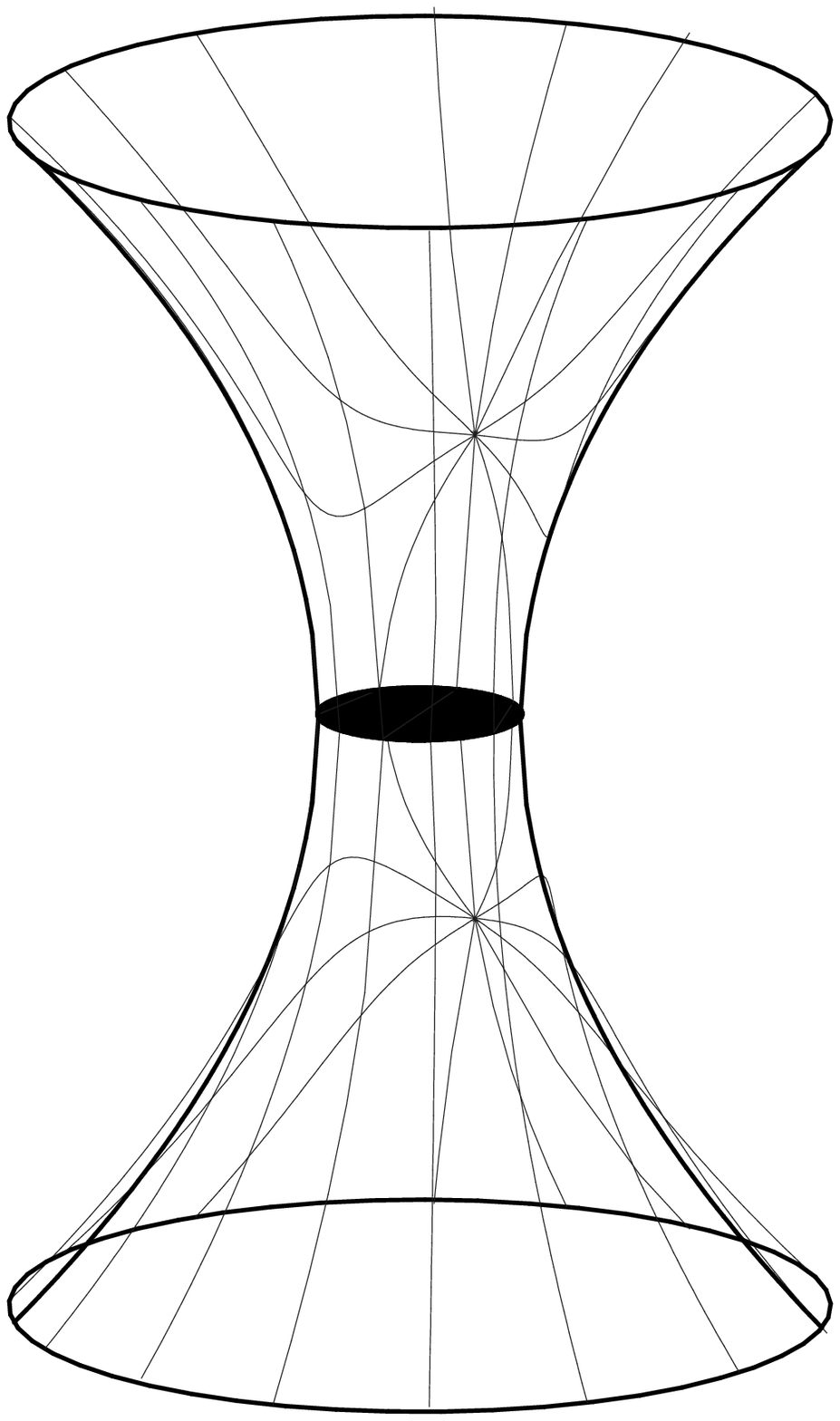}\\[0.4cm]
\mbox{(a)} & \mbox{(b)}\\
\end{array}$\\
\end{center}
\caption{Embedding diagram with the electric field lines of the particle shown in Fig.~\ref{fig:2}.
Fig.~(a) is taken from Wheeler's notebook.
}
\label{fig:2b}
\end{figure}

Later on, during the preparation of my volume with Rees and Wheeler, Johnny drew the electric lines of force as they should appear in an embedding diagram of the Schwarzschild solution (see Fig. \ref{fig:2b} (a)).
It is remarkable that he did this free hand, based on his great intuition. 
It is interesting to compare these same lines of force with those which have been recently recomputed \cite{bgr} by introducing the explicit computation of the embedding diagram (see Fig. \ref{fig:2b} (b)).
The exterior Schwarzschild solution can be visualized as a 2-dimensional hyperboloid embedded in the usual Euclidean 3-space, by suppressing the temporal and azimuthal dimensions associated with the symmetry.
The constant time equatorial slice has the reduced metric 
\begin{equation}
\label{embmetric}
ds^2=f_S(r)^{-1}dr^2+r^2d\phi^2\, .
\end{equation}
For $r>2{\mathcal M}$ the coordinate $r$ is spacelike, so this metric can be embedded in Euclidean space.
By employing regular cylindrical coordinates, the Euclidean metric is given by
\begin{equation}
\label{euclmetric}
ds^2=d\rho^2+\rho^2d\phi^2+dz^2\, ,
\end{equation}
with the same azimuthal angle $\phi$ for both metrics.
If we require that the metrics (\ref{embmetric}) and (\ref{euclmetric}) agree at constant $\phi$, we get the condition 
\begin{equation}
d\rho^2+dz^2=f_S(r)^{-1}dr^2\ .
\end{equation}
Setting $\rho=r$, this equation can be easily solved for $z$ as a function of $r$
\begin{equation}
\label{}
z = \int_{2{\mathcal M}}^{r}\left[\frac{2{\mathcal M}}{rf_S(r)}\right]^{1/2}dr
  = 2[2{\mathcal M}(r-2{\mathcal M})]^{1/2}\, ,
\end{equation}
with $z(2{\mathcal M})=0$.
Fig.~\ref{fig:2b} shows the embedding diagram with the electric field lines of the particle; it is in the curved space that the lines of force intersect the event horizon orthogonally (at which the field is strictly radial).


We note that also in this topic there is an open problem still to be resolved: we have also reconsidered \cite{bgr} the possibility of examining the problem not just of a point particle, but of a conducting sphere in the field of a Schwarzschild black hole, as done by Fermi in the case of a uniform gravitational field. 
This problem is not yet solved, although using important results obtained by Leaute and Linet \cite{leaute2}, we have been able to confirm the Fermi solution at least in the neighborhood of the test particle in Schwarzschild, where in a first approximation the gravitational field can be considered uniform.

\section{On the \lq\lq electric Meissner effect''}

My curiosity of how to justify the fact that Johnny did not succeeded in guessing a priori the lines of force near a Schwarzschild black hole still intrigued me a few years ago. I decided to look into the matter of a test particle near a Reissner-Nordstr\"om spacetime, motivated by results obtained in the mean time by Bicak and coworkers \cite{bicakpescara} for a magnetic dipole in the field of an extreme Reissner-Nordstr\"om and an extreme Kerr solution. 
My approach following the ``Wheeler theorem number 1'' was that in an extreme Reissner-Nordstr\"om solution with $Q={\mathcal M}$ no induced charge could exist, the reason being that any induced charge would make a part of the horizon surface overcritical and would generate a naked singularity. Instead of such catastrophic behaviour I was convinced that nature would have found the way to solve this paradox by not having lines of force crossing the horizon in the $Q={\mathcal M}$ case. 
I was also motivated in this thinking by an instant disagreement I felt reading a very publicized article by Parikh and Wilczek \cite{parikh}, where they simply extrapolated the results of a test particle near a Schwarzschild black hole to the case of a Reissner-Nordstr\"om metric without understanding the existence of this very profound underlying difference between the two cases. 

\begin{figure} 
\typeout{*** EPS figure 3}
\begin{center}
$\begin{array}{c@{\hspace{1in}}c}
\includegraphics[scale=0.33]{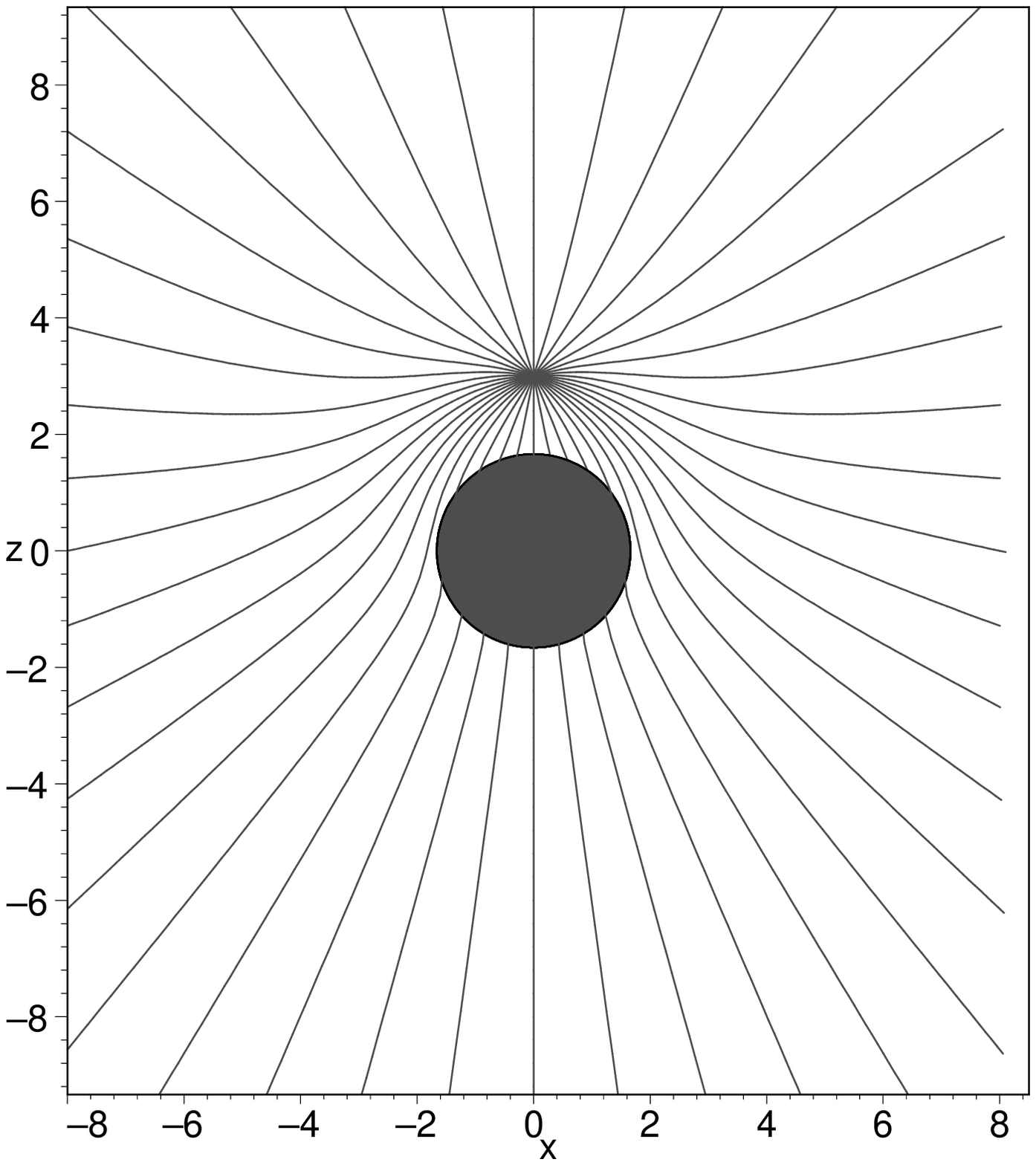}&
\includegraphics[scale=0.35]{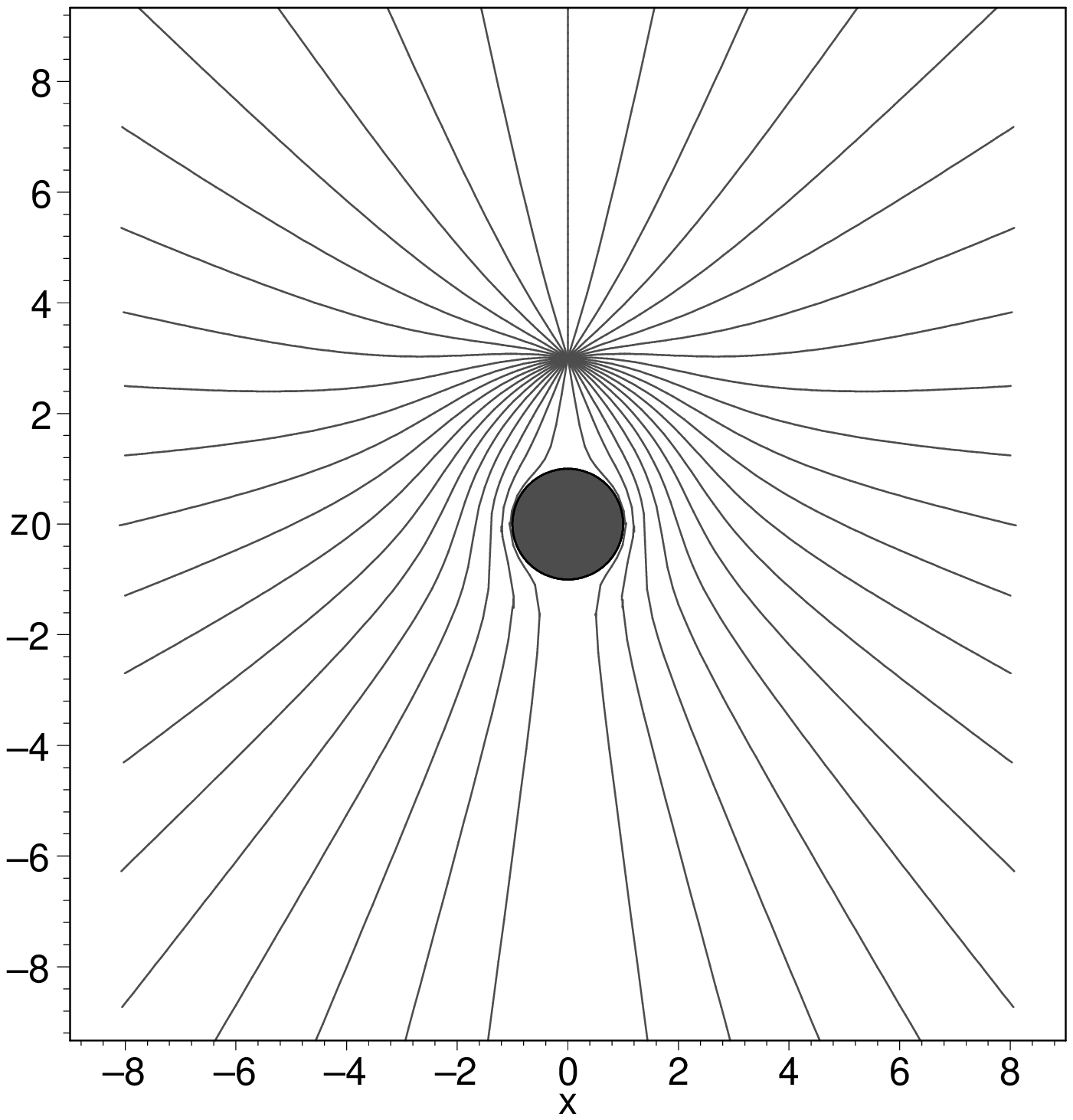}\\[0.4cm]
\mbox{(a)} & \mbox{(b)}\\
\end{array}$
\end{center}
\caption{Fig.~(a) shows the behavior of the lines of force of the test field alone with the charged particle at rest on the vertical axis $\theta=0$ at $r=b $ with $b/{\mathcal M} = 3$, for  $Q/{\mathcal M}=3/4$. 
Fig.~(b) corresponds to the extreme case, with the lines forced outside the (outer) horizon, the particle position being the same as in Fig.~(a).
The black circle represents the black hole horizon.}
\label{fig:4}
\end{figure}

I then proceeded with D. Bini and A. Geralico \cite{bgr} to study the set of equations for a test particle in a Reissner-Nordstr\"om spacetime
$$
\begin{array}{lc}
\label{RNmetric}
(\rm BGR1)\hspace{2.2cm}&  ds^2=- f(r)dt^2 + f(r)^{-1}dr^2+r^2(d\theta^2 +\sin ^2\theta d\phi^2)\, ,
\end{array}
$$
where $f(r)=1 - 2\mathcal{M}/r+Q^2/r^2$, with associated electromagnetic field
\begin{equation}
\label{RNemfield}
F_{\rm{RN}}=-\frac{Q}{r^2}dt\wedge dr\ .
\end{equation}
The horizon radii are $r_\pm={\mathcal M}\pm\sqrt{{\mathcal M}^2-Q^2}={\mathcal M}\pm\Gamma$.
Maxwell's equations (F2) reduce to the following equation for the electrostatic potential $V$ 
$$
\begin{array}{lc}
\label{poteq}
(\rm BGR5)\hspace{1.1cm}&  \displaystyle(r^2V_{,r})_{,r} +\frac{f(r)^{-1}}{\sin\theta}
\left[(\sin\theta V_{,\theta})_{,\theta} + \frac1{\sin\theta}(V_{,\phi})_{,\phi}\right]
=-4\pi r^2J^0\, .
\end{array}
$$
The solution corresponding to a charge $q$ placed at the point $r=b$ on the polar axis $\theta=0$ (with the same current density as (HR4)) has been derived by Leaute and Linet \cite{leaute} both as a multipole expansion analogous to (HR6) with
\begin{eqnarray}
f_l(r)&=&-\frac{(2l+1)!}{2^l(l+1)!l!\Gamma^{(l+1)}}\frac{(r-r_+)(r-r_-)}r 
      \frac{dQ_l}{dr}\left[\frac{r-\mathcal M}\Gamma\right]\qquad\,\,\,\, l=0,1,2, ...\nonumber\\
g_l(r)&=&\left\{
     \begin{array}{ll}
     1&\qquad\quad l=0\\
     \noalign{\medskip}\displaystyle\frac{2^ll!(l-1)!\Gamma^l}{(2l)!}\frac{(r-r_+)(r-r_-)}r 
     \frac{dP_l}{dr}\left[\frac{r-\mathcal M}\Gamma\right]&\qquad\quad l=1,2, ... 
     \end{array}
     \right. 
\end{eqnarray}
and in the closed form  
\begin{equation}
\label{solRNpot}
V = \frac q{b r} \frac{(r-{\mathcal M})(b-{\mathcal M})
 -\Gamma^2\cos\theta}{D_{\rm{RN}}} + \frac{q{\mathcal M}}{b r}\, ,
\end{equation}
with
\begin{equation}
\label{denRN}
D_{\rm{RN}}
= [(r-{\mathcal M})^2+(b-{\mathcal M})^2 
- 2(r -{\mathcal M})(b-{\mathcal M})\cos\theta 
- \Gamma^2\sin^2\theta]^{1/2}\, . 
\end{equation}
We also generalized to the Reissner-Nordstr\"om case the discussion of the lines of force and associated properties of the horizon presented above for the Schwarzschild case. 
The induced charge density on the horizon is easily evaluated
$$
\begin{array}{lc}
\label{sigmaRN}
(\rm BGR7)\hspace{2.6cm}& \sigma_{\rm{RN}}^H(\theta)
=\displaystyle\frac{q}{4\pi b} \frac{[\Gamma(1+\cos^2\theta)-2(b-\mathcal M)\cos\theta]\Gamma}
                  {[b-\mathcal M-\Gamma\cos\theta]^2(\Gamma+\mathcal M)}\, . 
\end{array}
$$ 
Note that $\sigma_{\rm{RN}}^H(\theta)$ becomes identically zero in the extremely charged case where $\Gamma=0$. 
As the hole becomes extreme, an effect analogous to the ``magnetic Meissner effect'' in the presence of superconductors arises for the electric field, with the electric field lines of the test charge being forced outside the outer horizon (see Fig.~\ref{fig:4}). But this time the effect is not on a magnetic field, but is on the electric field, and it is not due to a superconducting sphere, but to the spacetime around an extreme Reissner-Nordstr\"om black hole.


\section{On Zerilli's solution}

It must be emphasized that this is only a preliminary result. This behavior must be confirmed by integrating the more general set of equations describing the full Einstein-Maxwell perturbation equations introduced by Zerilli \cite{zerilli}
\begin{eqnarray}
\label{EinMaxeqs}
\tilde G_{\mu \nu }&=&8\pi \left( T_{\mu \nu }^{\rm part} + \tilde T_{\mu \nu }\right) ,\nonumber\\
\label{EM}
\tilde F^{\mu \nu }{}_{;\,\nu }&=& 4\pi  j^{\mu }_{\rm part} , \quad {}^* \tilde F^{\alpha\beta}{}_{;\beta}=0\ ,
\end{eqnarray}
where the quantities denoted by the tilde refer to the total electromagnetic and gravitational fields, at the first order of the perturbations, i.e.
\begin{eqnarray}
\label{pertrelations}
\tilde g_{\mu \nu }&=&g_{\mu \nu } + h_{\mu \nu }\, ,\nonumber\\
\tilde F_{\mu \nu }&=&F_{\mu \nu }+  f_{\mu \nu }\, ,\nonumber\\
\tilde T_{\mu \nu }&=& \frac1{4\pi}\left[\tilde g^{\rho \sigma }\tilde F_{\rho \mu }\tilde F_{\sigma \nu }
    - \frac14\tilde g_{\mu \nu }\tilde F_{\rho \sigma }\tilde F^{\rho \sigma }\right]\, ,\nonumber\\
\tilde G_{\mu \nu }&=&\tilde R_{\mu \nu }-\frac12\tilde g_{\mu \nu }\tilde R\, ,
\end{eqnarray}
where $T_{\mu \nu }^{\rm part} $ and $j^{\mu }_{\rm part} $ are respectively the stress-energy tensor and the $4$-current associated with a particle of mass $m$ and charge $q$. 
The corresponding quantities without the tilde refer to the background Reissner-Nordstr\"om metric (BGR1) and its associated electromagnetic field (\ref{RNemfield}).

For a point charge of mass $m$ and charge $q$ at rest at the point $r=b $ on the polar axis $\theta=0$, the only nonvanishing components of the stress-energy tensor and of the current density are given by
\begin{eqnarray}
\label{sorgenti}
j^{{0}}_{\rm{part}}&=&{\frac {q}{2\pi {b}^{2}}}\delta  \left( r-b \right) \delta  \left( \cos  \theta -1 \right)
\, ,\nonumber\\
T_{{00}}^{\rm{part}}&=&{\frac {m}{2\pi {b}^{2}}}f(b)^{3/2}\delta  \left( r-b \right) \delta  \left( \cos \theta -1 \right)\, .
\end{eqnarray}
The perturbation equations are obtained from the system (\ref{EM}), keeping terms to first order. 
That has been done \cite{bgr} and the existence of the ``electric Meissner effect'' has been confirmed. There are also some very exciting new results on the two-body solution in a Reissner-Nordstr\"om geometry, which we will discuss in the near future.

\section{Conclusions}

The examples we have given well illustrate the caution we should apply in stating that black holes behaves as perfect conductors and a special care should be used in establishing correspondence and analogies between classical physics and general relativistic regimes. See e.g. the statement I did in the book in honor of the festschrift of Hagen Kleinert \cite{kleinert}: ``The analogies between classical regimes and general relativistic regimes have been at times helpful in giving the opportunity to glance on the enormous richness of the new physical processes contained in Einstein's theory of space-time structure. In some cases they have allowed to reach new knowledge and formalize new physical laws [...]. Such analogies have also dramatically evidenced the enormous differences in depth and physical complexity between the classical physics and general relativistic effects. The case of extraction of rotational energy from a neutron star and a rotating black hole are a good example. In no way an analogy based on classical physics can be enforced on general relativistic regimes. Such an analogy is too constraining and the relativistic theory shows systematically a wealth of novel physical circumstances and conceptual subtleties, unreachable within a classical theory. The analogies in the classical electrodynamics we just outlined are good examples.'' 

It is very interesting that the combined Einstein-Maxwell equations still offer new challenges leading to unexplored physical phenomena. These results offer the possibility of reaching a better understanding of the solutions of both the Einstein and Einstein-Maxwell equations.
In both these topics M.me Choquet-Bruhat has made profound contributions and it is with great pleasure that I present these results to her in honor of her eightieth birthday. I am also very happy to share this celebration by recalling two very good friends of ours, Zel'dovich and Wheeler, both companions with us in the search for a deeper meaning of Einstein's great theory. Last, but not least, after all Johnny was right!

\end{document}